\documentclass{pasj00}

\usepackage{graphicx}
\usepackage{tabularx}
\usepackage{here}
\draft

\begin{document}
\SetRunningHead{M.Ohno et al.}{Suzaku-WAM and Konus-Wind Observation of Short Gamma-Ray Bursts}
\Received{2007/01/29}
\Accepted{2007/03/07}

\title{\bf Spectral Properties of Prompt Emission of Four Short Gamma-Ray Bursts Observed by the
 Suzaku-WAM and the Konus-Wind}

\author{Masanori \textsc{Ohno}$^1$, Yasushi \textsc{Fukazawa}$^1$ Takuya
\textsc{Takahashi}$^1$,  \\Kazutaka
\textsc{Yamaoka}$^2$, Satoshi \textsc{Sugita}$^2$,  Valentin \textsc{Pal'shin}$^3$,
Takanori \textsc{Sakamoto}$^4$,  \\ Goro \textsc{Sato}$^4$,  Kevin
\textsc{Hurley}$^5$, 
Dmitry \textsc{Frederiks}$^3$, Philipp \textsc{Oleynik}$^3$, Mikhail \textsc{Ulanov}$^3$ \\
Makoto \textsc{Tashiro}$^6$,  Yuji \textsc{Urata}$^6$, Kaori
\textsc{Onda}$^6$,  Toru \textsc{Tamagawa}$^7$, \\ Yukikatsu \textsc{Terada}$^7$, Motoko
\textsc{Suzuki}$^7$, Hong \textsc{Soojing$^8$}}%

\affil{$^1$Department of Physical Sciences, School of Science, Hiroshima
 University \\ 1-3-1 Kagamiyama, Higashi-Hiroshima, Hiroshima 739-8526}
\email{ohno@hirax7.hepl.hiroshima-u.ac.jp}

\affil{$^2$Department of Physics and Mathematics,  Aoyama Gakuin University,\\ 5-10-1,Fuchinobe, Sagamihara 229-8558}

\affil{$^3$Ioffe Physico-Technical Institute, Laboratory for Experimental Astrophysics, \\ 26 Polytekhnicheskaya, St. Petersburg 194021, Russa}

\affil{$^4$NASA Goddard Space Flight Center, Greenbelt, MD 20771}

\affil{$^5$University of California, Berkeley, Space Sciences Laboratory, \\
7 Gauss Way, Berkeley, CA 94720-7450}

\affil{$^6$Department of Physics, Saitama University, \\255 Shimo-Ohkubo, Sakura, Saitama 338-8570}

\affil{$^7$Institute of Physical and Chemical Research (RIKEN), \\ 2-1 Hirosawa, Wako, Saitama 351-0198}

\affil{$^8$Laboratory of Physics College of Science and Technology,
Nihon University \\ 7-24-1 Narashinodai, Funabashi, Chiba 274-751}


\maketitle

\begin{abstract}

We have performed a joint analysis of prompt emission from four
 bright short gamma-ray bursts (GRBs) 
with the Suzaku-WAM and the Konus-Wind experiments. 
This joint analysis allows us to investigate the spectral
 properties of short-duration bursts over a wider energy band with a higher
 accuracy. 
We find that these bursts have a high E$_{\rm peak}$, around 1 MeV
 and have a harder power-law component than that of long GRBs. 
However, we can not determine whether these spectra follow the cut-off 
power-law model or the Band model.

We also investigated the spectral lag, hardness ratio, inferred isotropic radiation
 energy and existence of a soft emission hump, in order to classify 
them into short or long GRBs  using several criteria, in addition to
 the burst duration. 
We find that all criteria, except for the existence of the soft hump,
 support the fact that our four GRB samples are correctly classified as
 belonging to the short class. In addition, our broad-band analysis
 revealed that 
there is no evidence of GRBs
with a very large hardness ratio, as seen in the BATSE short 
GRB sample, and that the spectral lag of our four short GRBs is consistent
 with zero, even in the MeV energy band, unlike long GRBs. 
 Although our short GRB samples are still limited, these results suggest
   that the spectral hardness of short GRBs might not  
  differ significantly from that of long GRBs, 
 and also that the spectral lag at high energies could be a strong criterion for burst classification.

\end{abstract}

\section{Introduction}

The bimodal distribution of the duration of gamma-ray bursts (GRBs)
indicates that there are two distinct classes of events. The 
long-duration bursts have typical durations of around 20 s, while
the short-duration bursts (about 1/4 of the total) have durations 
of around 0.3 s
(\cite{Mazets 1981}; \cite{Norris 1984}; \cite{Dezalay 1992}; 
\cite{Hurley 1992}; \cite{Kouveliotou 1993}; \cite{Norris 2000}).
 This suggests that the short and long GRBs may have different 
progenitors. 
Core collapse of massive stars and
mergers of  compact binaries are considered likely models for the
long and short GRBs, respectively (\cite{Katz_Canel}; \cite{Ruffert
1999}). 
Recently, thanks to the 
rapid position information provided by HETE-2 (\cite{Rikker 2003}) and
Swift (\cite{Gehrels 2004}),  afterglow 
observations have progressed dramatically and 
 X-ray and optical afterglow
emissions were discovered from some short GRBs (\cite{Hjorth 2005};
\cite{Fox 2005}), as well as long ones. Some short GRB afterglows have
been found to be  associated with galaxies not undergoing star formation,
while some long GRBs have been found to be associated with
energetic supernovae. 
These results support the hypothesis that long and short GRBs indeed
have different progenitors.

From the point of view of prompt gamma-ray emission, although the
spectral characteristics of long GRBs have been well studied
(\cite{Frontera 2000}: BeppoSAX; \cite{Kaneko 2006}: BATSE;
\cite{Sakamoto 2005b}: HETE-2), 
our understanding of that of short GRBs is still
incomplete, in part due to their very short durations. 
The spectral
characteristics of BATSE short GRBs are often characterized by the
hardness ratio, that is, the ratio of 100--300 keV to 25--100 keV
counts or fluence
(\cite{Kouveliotou 1993}; \cite{Cline 1999}).
These studies suggest that, although there is considerable overlap,  short GRBs
tend to be harder than long GRBs.  
Paciesas et al. (2001) and Ghirlanda
et al. (2004) compared the spectral parameters of bright BATSE short GRBs with
those of long GRBs by spectral fitting. 
They pointed out that the
spectra of short GRBs are well described by a cut-off power-law model
and that the Band model (\cite{Band 1993}) did not improve the fit. 
They also confirmed that the spectra of short
GRBs are harder than those of long GRBs. This was found to be a
consequence of a flat low energy photon index, rather than
a difference in the peak energies. 

There is another complicating issue in the classification of short and 
long GRBs.
Since their duration distributions overlap, other distinguishing criteria 
have been proposed, such as differences in the host galaxy (\cite{Hjorth
2003}; \cite{Hjorth 2005}), spectral hardness (\cite{Cline 1999}), 
spectral lag (\cite{Norris 2002}; \cite{Norris 2006}), 
isotropic radiation energy (\cite{Amati 2002}; \cite{Amati 2006}), 
and the existence of a soft hump (\cite{Norris 2006}), or a combination of
these and other characteristics (\cite{Donaghy 2006}). 
Indeed, some GRBs cannot be classified as short or long
 using the burst duration alone. For example, GRB 040429
 was classified as short using the burst duration (2.39 s), 
but other properties such as spectral lag and hardness resembled
those of long GRBs (\cite{Wiersema 2004}; \cite{Fox_Moon 2004}; \cite{Donaghy 2006}).
GRB 051227 (\cite{Hullinger 2005}; \cite{Barthelmy 2005a};
\cite{Sakamoto 2005})
had a long enough duration of 8.0 s to be classified as long. 
However, this burst had no significant spectral lag in the
initial spike, which is seen in many long GRBs (\cite{Norris 1996}), and
it also had a long soft hump in the light curve. These two properties
are similar to those of short GRBs.
GRB 060614 (\cite{Gehrels 2006})
exhibited a long duration (102 s). However, it did not have any 
significant spectral lag in the initial spike, and
also there were no signs of any associated supernova, despite its
distance of z = 0.125, which should have been close enough
to detect it. These examples show why the burst duration should no
longer be considered a sole indicator for distinguishing between the
short and long classes of bursts. Indeed Donaghy et al. (2006) have
suggested the terms ``short population bursts'' and ``long population
bursts'' to distinguish the two classes. In this paper, we will use
simply ``short'' and ``long'' to describe these classes.

Here, we report on the joint spectral analysis of
four bright short GRBs 
simultaneously observed by the Suzaku-WAM (Suzaku Wide-band All-sky
Monitor) and Konus-Wind. 
The WAM and  Konus have wide energy ranges, from 50--5000 keV and 10--10000
keV, respectively, and the WAM has the largest effective area from 300
keV to 5000 keV of any current experiment with spectral capabilities.
Thus, we can investigate the spectral characteristics of 
 short GRBs up to MeV energies.
We also classify the four GRBs as short or long using our data with
high statistics up to the MeV energy region using a number of criteria: 
spectral lag, hardness ratio, isotropic radiation energy and existence
of a soft hump.

\section{Instruments and Observations}

\subsection{Instruments}

\subsubsection{Suzaku-WAM}

The Suzaku Wide-band All-sky Monitor (WAM) is the active shield of the Hard X-ray detector
(HXD-II) (\cite{Takahashi 2006}; \cite{Kokubun 2006})
aboard Suzaku (\cite{Mitsuda 2006}). 
It consists of large-area, thick BGO crystals,  
and  is also designed to monitor the entire sky from 50 keV to 5 MeV
with a large effective area. The large effective area from 300 keV to 5
MeV (400 cm$^2$ even at 1 MeV) surpasses those of other currently
operating experiments with spectral capability, and enables us to perform 
wide-band spectroscopy of GRBs with high sensitivity (\cite{Yamaoka
2005}; \cite{Yamaoka 2006}). The WAM is subdivided into four
detectors, numbered  WAM0 to WAM3, located at each side of 
the main detector of the
HXD-II. We utilize the azimuthal ($\phi$ [degree]) and 
zenith ($\theta$ [degree]) angle to determine 
the incident direction of GRBs. The azimuthal zero angle ($\phi=0$) 
is defined to be perpendicular to the WAM1 detector plane, and it
increases in the order WAM0, WAM3, and WAM2. 
The WAM0 detector faces the solar paddle of the satellite and it
always views the Solar direction.
 The polar axis that defines the zenith angle $\theta$ is toward the
 Suzaku field-of-view of the main detectors.
Thus the on-axis direction of each WAM detector correspond to
($\theta$, $\phi$) = (90, 0) for WAM1, (90, 90) for WAM0,
(90, 180) for WAM3, and  (90, 270)
for WAM2.
 
The WAM outputs two data types, the transient (TRN) data
and the gamma-ray burst (GRB) data. The TRN data are always accumulated
with 1 s time resolution and 55 energy channels. This can be used to
monitor the bright soft gamma-ray sources with the Earth
occultation method, as was done by  CGRO/BATSE (\cite{Ling 2000}).
On the other hand, the GRB data are recorded only for 64 s when
the GRB trigger is activated, and the data cover 8 s before and 56
s after the trigger time. The GRB data have four
energy channels with 1/64 s time resolution, in addition
to the spectral data in 55 pulse height channels with 0.5 s time
resolution. This allows us to
perform both spectroscopy and timing analysis using the GRB
data.
In the initial phase before 2006 March 20, these two time resolutions
for GRB data were set to 1/32 s and 1.0 s, respectively.

The response matrix of the WAM is very complicated, depending on both
the azimuthal and zenith angles of GRBs, because the WAM is inside 
the satellite body and suffers heavy
absorption from it. 
In order to calculate the detector response of the WAM, we utilize a Monte
Carlo simulation based on the Geant4 code, including many kinds of 
detector characteristics. In the pre-flight calibration, we measured
these detector characteristics and developed a WAM response generator
(\cite{Ohno 2005}). From the results of this pre-flight calibration, 
we found that the response generator can reproduce the
measured spectral shape, but the measured effective area 
of the WAM varies drastically, depending
on the gamma-ray incident angle. 
The trend of this angular response is different among
the four WAM detectors. The angular responses of 
the WAM1 and WAM3 detectors is complicated, and this trend is similar 
in both cases. We believe that this is caused by absorption from 
numerous electronics devices attached inside the satellite panels. 
The WAM2 detector is the most heavily affected, since the Ne/He chamber for
the XRS is located inside the 
WAM2 field-of-view. 
This is not the case for the angular
response of the WAM0 detector, 
and it can be approximately described
by a simple cosine function. This is because only the solar paddle is
attached to the satellite panel, facing the WAM0 detector. 

In order to estimate the uncertainty in the angular response,
we performed an in-flight cross-calibration between the WAM, Konus, and the
Swift/BAT, using the data of simultaneously detected GRBs,
 and found that the measured spectral shapes agree quite well
 up to the MeV energy region, but
that the observed flux fluctuates typically by 10--20\%  and by 40\% at
most above 100 keV. This uncertainty is consistent with pre-launch
expectations. Below 100 keV, there are still larger flux
uncertainties of about 50\%. We also confirmed that the WAM0 detector
has the most reliable detector response, with a flux uncertainty of 20\% 
(Sakamoto et al. 2007, in preparation).

\subsubsection{Konus-Wind}


The Konus-Wind instrument (Aptekar et al. 1995) is a gamma-ray 
spectrometer aboard the GGS-Wind spacecraft. It consists of two 
identical detectors, S1 and S2, which observe correspondingly the south 
and north ecliptic hemispheres in an all-sky monitoring mode. Each 
detector is a cylindrical NaI(Tl) crystal 13 cm in diameter and 7.5 cm 
in height. The experiment operates in triggered and waiting modes. In 
the triggered mode the burst time histories are recorded in three energy 
windows (G1, G2, and G3) 0.512~s before and 229.632~s after the trigger,
with a variable time resolution from 2~ms up to 256~ms.
64 energy spectra are measured in two partially overlapping energy 
ranges, nominally 10--750 keV and 0.2--10 MeV. In each range they are 
recorded in 63 channels with time resolutions starting at 64 ms, and 
subsequent spectra have resolutions from 256 ms to 8.192 s. In the 
waiting mode, each detector measures the count rate with a resolution of 
2.944~s in three energy windows.  For the GRBs considered here, the 
actual spectral ranges were 21 keV--16 MeV for the S1 detector and 18 keV--14 MeV for the S2 detector; the actual time history windows were 21--83, 
83--360, 360--1360 keV (S1), and  18--70, 70--300, 300--1160 keV (S2).
The detector response matrices depend only on the
zenith ($\theta$) angle because of its axi-symmetric structure.

\subsubsection{The Interplanetary Network}

The Interplanetary Network (IPN) is a group of spacecraft with
gamma-ray burst detectors.  Its main purpose is to monitor the
entire sky for gamma-ray bursts with close to a 100\% duty cycle,
and to localize bursts with up to arcminute accuracy.  This is done by comparing the arrival times of a GRB
among the spacecraft.
Since the third IPN began operations in 1990, over 25 spacecraft
have participated in it, including Konus-Wind (\cite{Hurley
2006a}).
The Suzaku-WAM
is the latest experiment to join the IPN.
It now plays an important role in it, and many GRBs have already been
localized using WAM data (\cite{Golenetskii 2006a}; \cite{Cumming
2006}; \cite{Hurley 2006b}; \cite{Golenetskii 2006b}).

When a burst is detected by three widely separated spacecraft, a
very accurate localization can usually be obtained.  For the bursts
discussed here, however, the Ulysses GRB experiment was off, and Mars
Odyssey was the only interplanetary spacecraft.  The resulting error
boxes in this case tend to be large.
Table \ref{ipntable} shows which spacecraft observed the events discussed
in this paper.  Note that GRB 060317 was not observed by Mars Odyssey,
and therefore the uncertainty in its localization is the largest.
Table \ref{ipntable2} gives the final coordinates of the error boxes.
In most cases, the error boxes were not small enough and/or determined
rapidly enough to warrant follow-up searches.  
In the case of GRB 060429,
however, GCN Circulars were issued (\cite{Hurley 2006b}; \cite{Golenetskii 2006b}),
and a Swift target of opportunity observation was carried out.  However,
for this and the other events, no counterparts were identified at any
wavelength.


\subsection{The Short GRB Sample}

More than 70 GRBs triggered the WAM and Konus simultaneously
between August 2005 to October 2006. In order to perform joint spectral
analysis of short GRBs, we selected well
localized short GRBs and constructed detector response matrices.
Only 8 short GRBs were localized well enough by Swift and/or the IPN 
in this period, and 4 of them (GRB 051127, GRB 060317, GRB 060429, 
and GRB 060610) had hard enough spectra to detect in the highest 
energy band light curves of both instruments. 
Therefore, we select them to
investigate the spectral properties of short GRBs up to the MeV energy region.

GRB 051127 triggered the WAM and Konus at T$_0$(WAM) = 22:55:19.896 UT,
and T$_0$(KW) = 22:55:15.860 UT 2005 November 27. The incidence angles 
on the detectors could be determined thanks to the 
IPN localization  (\cite{Hurley 2005}). The incident direction of this
GRB was the nearest to on-axis for the WAM0 detector of the four, 
and thus the uncertainty in the detector response is the lowest for the WAM.
GRB 060317 was very bright  and it triggered 
the WAM and Konus at T$_0$(WAM) = 11:17:39.104 UT, and T$_0$(KW) = 11:17:35.996
 UT 2006 March 17. The incidence  angle on the WAM
was also good for the WAM0 detector.
At T$_0$(WAM) = 12:19:51.031 UT, and T$_0$(KW) = 12:19:49.712 UT 2006
April 29, 
the WAM and Konus detected the bright short GRB 060429.  
GRB 060610 triggered the WAM1 detector  
and the S2 detector of Konus 
at T$_0$(WAM) = 11:22:23.544 UT and  T$_0$(KW) = 11:22:22.632 UT 2006 June 10.
For this GRB, the incident direction was closer
to on-axis for the WAM1 detector than WAM0. But
this angle was determined to be good in the cross-calibration, 
and the flux uncertainty still should be within 20\%.
The detections are summarized in table \ref{ipntable2}.

\section{Data Analysis}

We performed spectral analysis by applying three spectral models.
The first one is a simple
power-law (PL) model,

\begin{equation}
N(E)=A \times \Biggl(\frac{E}{100}\Biggr)^{\alpha},
\end{equation}
where  A is the
normalization constant at 100 keV in photons cm$^{-2}$ s$^{-1}$ keV $^{-1}$, and $\alpha$ is the power-law photon
index.

The second model is a power-law with an exponential cutoff (CPL)
model,

\begin{equation}
N(E)=A \times \Biggl(\frac{E}{100}\Biggr)^{\alpha} {\rm exp}\Biggl(-\frac{E(2+\alpha)}{E_{\rm peak}}\Biggr),
\end{equation}
where E$_{\rm peak}$ is the peak energy in the $\nu$F$_\nu$ spectrum, and
 represents the energy at which most of the power is emitted.
 
The third model is a smoothly connected broken power-law model known as 
the Band model (\cite{Band 1993}):

\begin{eqnarray}
N(E) & = & A \times \Biggl(\frac{E}{100}\Biggr)^{\alpha} {\rm exp}
 \Biggl(-\frac{E(2+\alpha)}{E_{\rm peak}}\Biggr),
 ~~~~~~~~~~~~~~~~~~~{\rm for}~ E <
 \frac{(\alpha-\beta)E_{\rm peak}}{(2+\alpha)}\nonumber\\
     &   & A \times \Biggl(\frac{E}{100}\Biggr)^{\beta} \Biggl[\frac{(\alpha - \beta) E_{\rm
      peak}}{100(2+\alpha)}\Biggr]^{(\alpha-\beta)} {\rm exp}(\beta -
      \alpha),  ~{\rm for}~ E \geq
 \frac{(\alpha-\beta)E_{\rm peak}}{(2+\alpha)},
   \end{eqnarray}
where $\alpha$ is the power-law photon index in the lower energy band,
and $\beta$ is the index in the higher energy band.

First, we performed 
fits for each instrument alone. After confirming the consistency
between the WAM and Konus, we performed a joint fit. 
Selecting the time regions was
important for this analysis because the WAM
and Konus have different time resolutions for the spectral data. 
As  mentioned in section 2, the WAM has a 1.0 or 0.5 s
 time resolution for
spectral data, and can accumulate spectra both before and after the
GRB trigger. On the other hand, Konus can accumulate 
spectra only  after the trigger, and the 
time resolution varies depending on the burst intensity. 
In the case of the four short GRBs analyzed here, the
time resolution of the Konus spectra is 64 ms for first four intervals and
8.192 s for remainder. 
Therefore, the time regions cannot be completely identical for the
WAM and Konus, and
we have selected the times for each instrument that are
as close as possible
between the two.  To account for the differences in the accumulation times, 
we introduced a constant factor in the joint fits.  
The background spectra were
extracted from both before and after the WAM time regions, 
but only after the time region for
Konus. 
As mentioned in subsubsection 2.1.1, the detector response of the WAM strongly
depends on the position of the GRB, and thus the error box of the IPN
localization might cause some uncertainties for spectral analysis of the
WAM. In order to estimate this effect, we calculated the detector
response of the WAM for all corners of the box, as shown
in table $\ref{ipntable2}$, and performed spectral analysis
by using each of them. This uncertainty
is included as the systematic error.
We used XSPEC version 11.3.2 for spectral analysis
(\cite{Arnaud 1996}). Throughout this paper, the quoted uncertainties
are given at the 90\% confidence level for one interesting parameter.
A summary of the fitting results is given in table
\ref{fit_summary_tbl}. 

We also examined the spectral lag using the cross-correlation function
(CCF) between the light curves in
two energy bands (\cite{Norris 2002}; \cite{Norris 2006}; \cite{Link
1993}; \cite{Band 1997}; \cite{Yi 2006}). 
If $v_1$(t) and $v_2$(t) are the light
curves in two different energy bands, the CCF is defined by 
\begin{equation} 
{\rm CCF}(\tau_{lag}) = \frac{\sum_{t} v_1(t)v_2(t+\tau_{lag})}{N
  \sqrt{\sigma_{v_1} \sigma_{v_2}}}~~~~~    ,{\rm where}~~~ \sigma^2_v = \frac{1}{N}\sum_{i=1}^N v^2_i,\\
\end{equation}
where N is the number of data points in the light curve.
We used four energy bands for the WAM light curves: 
50--110 keV, 110--240 keV, 240--520 keV, and  520--5000 keV. 
After calculating the CCF as a function
of lag $\tau_{lag}$, we obtained the peak value of $\tau_{lag}$ 
by fitting it with a Gaussian profile (\cite{Yi 2006}). In this formula, positive
spectral lag means that the spectrum has hard to soft evolution.

\subsection{GRB 051127}

\begin{figure}[htbp]
\begin{minipage}{8cm}
\begin{center}
\rotatebox{-90}{\resizebox{!}{8cm}{\includegraphics{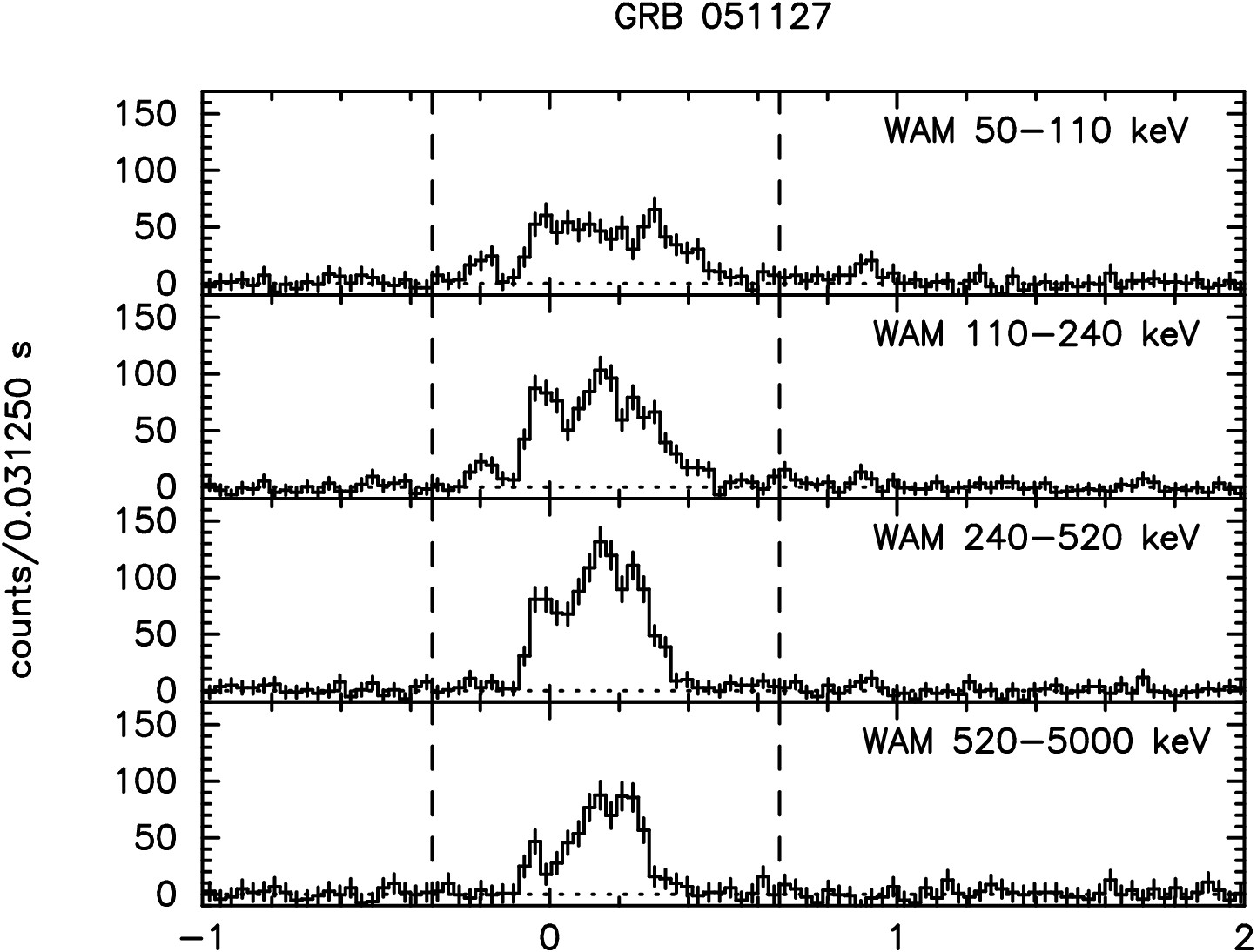}}}
\rotatebox{-90}{\resizebox{!}{8cm}{\includegraphics{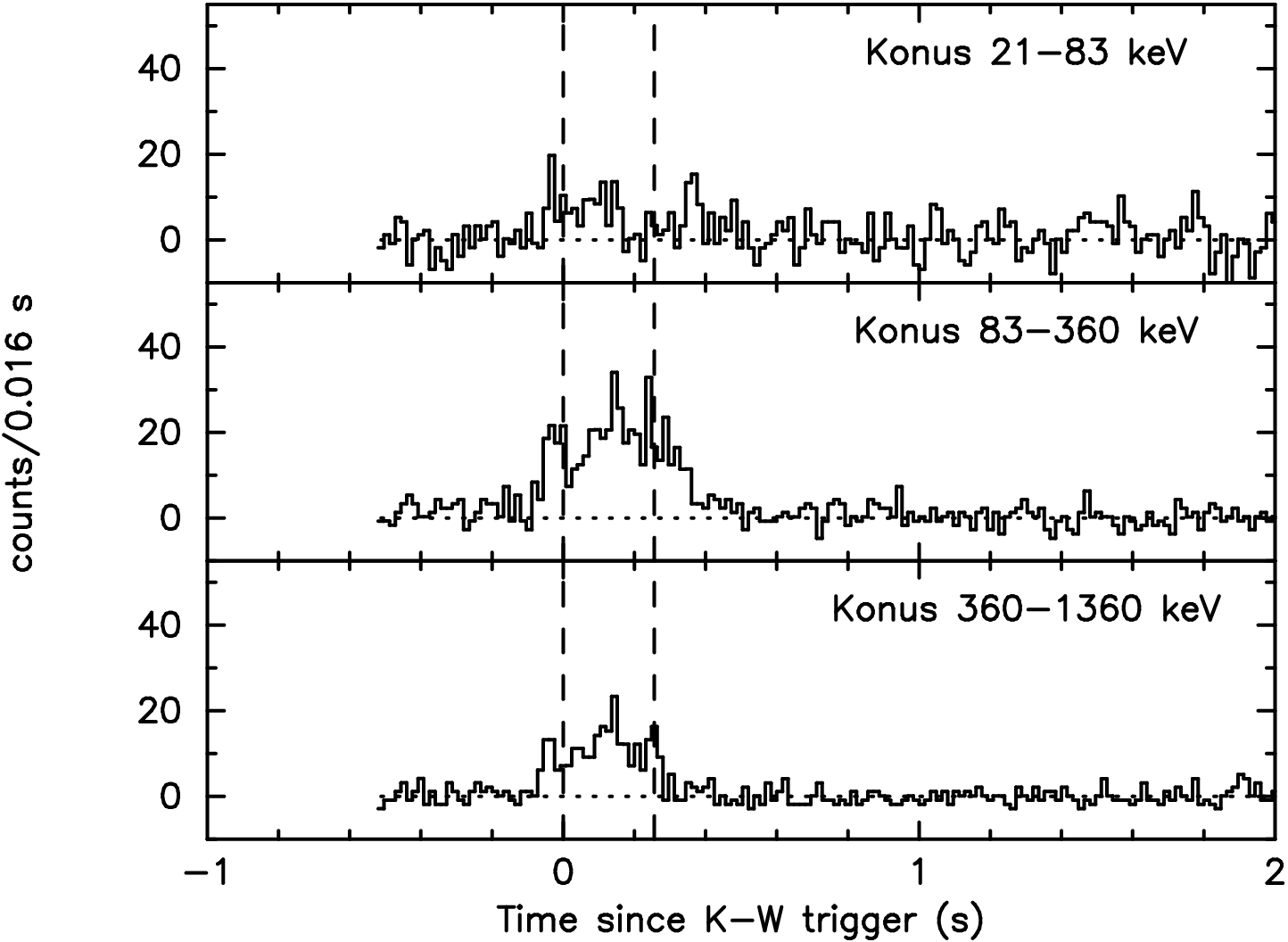}}}
\end{center}
\begin{center}
\rotatebox{-90}{\resizebox{!}{8cm}{\includegraphics{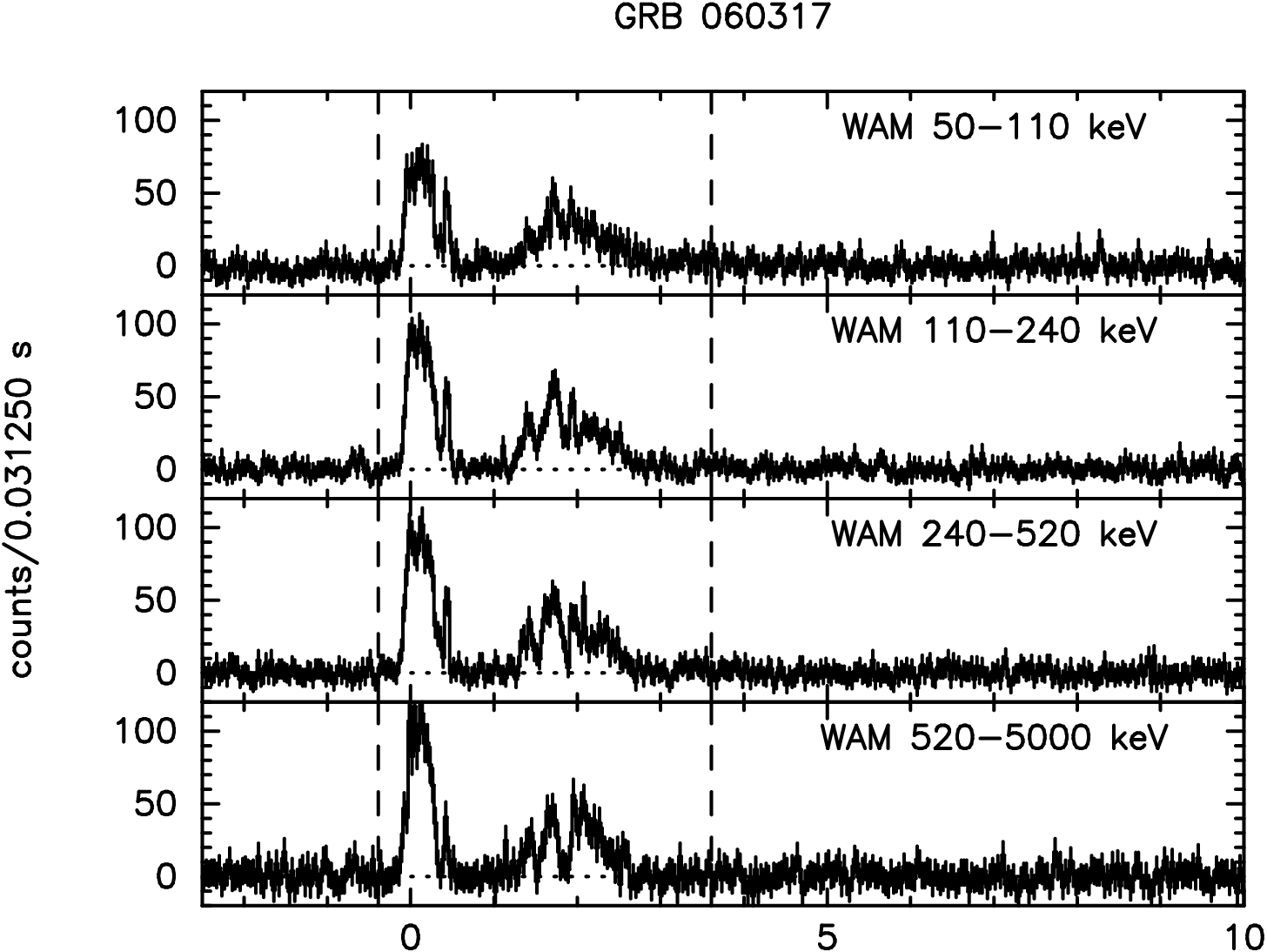}}}
\rotatebox{-90}{\resizebox{!}{8cm}{\includegraphics{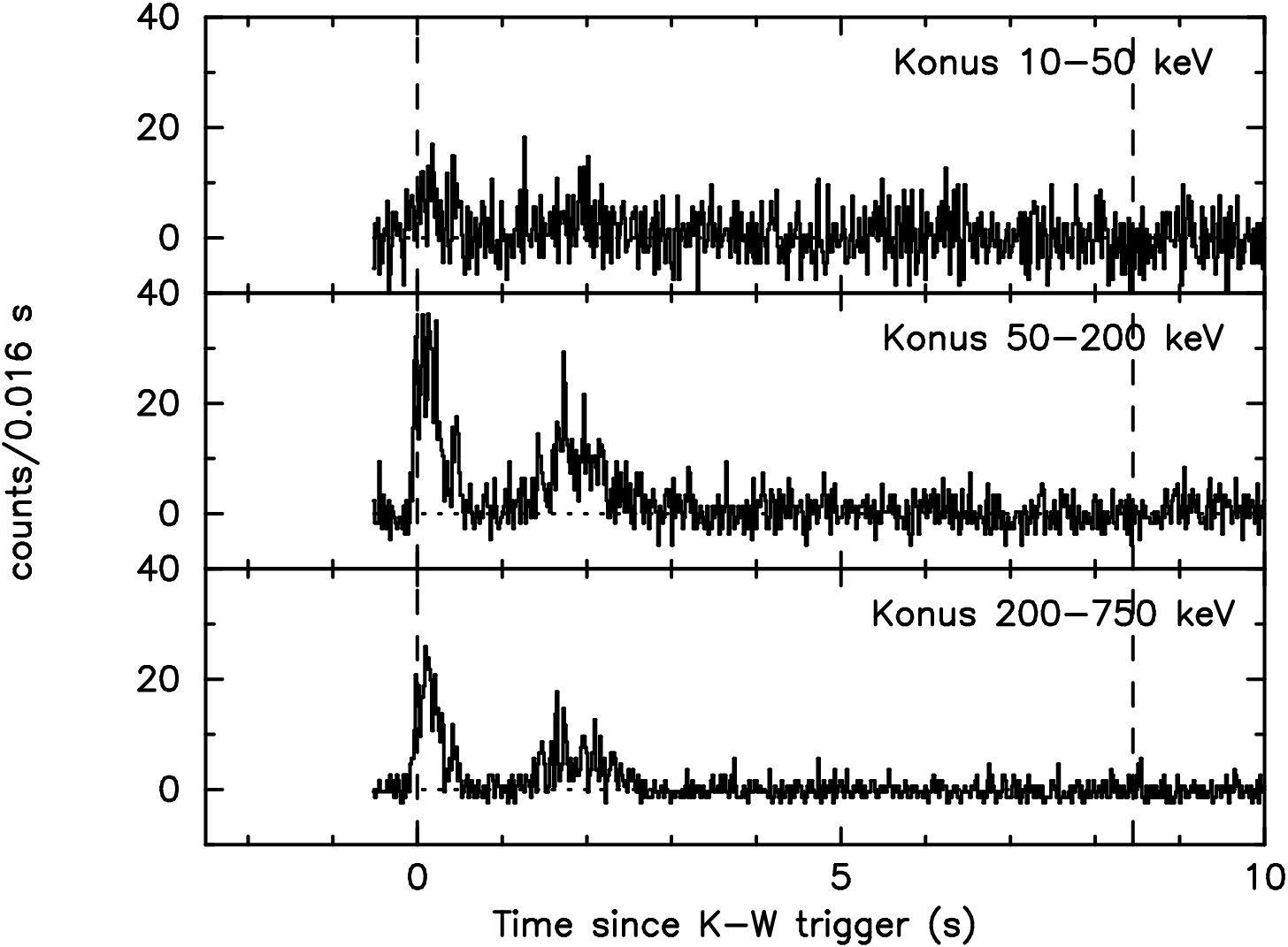}}}
\end{center}
\end{minipage}
\begin{minipage}{8cm}
\begin{center}
\rotatebox{-90}{\resizebox{!}{8cm}{\includegraphics{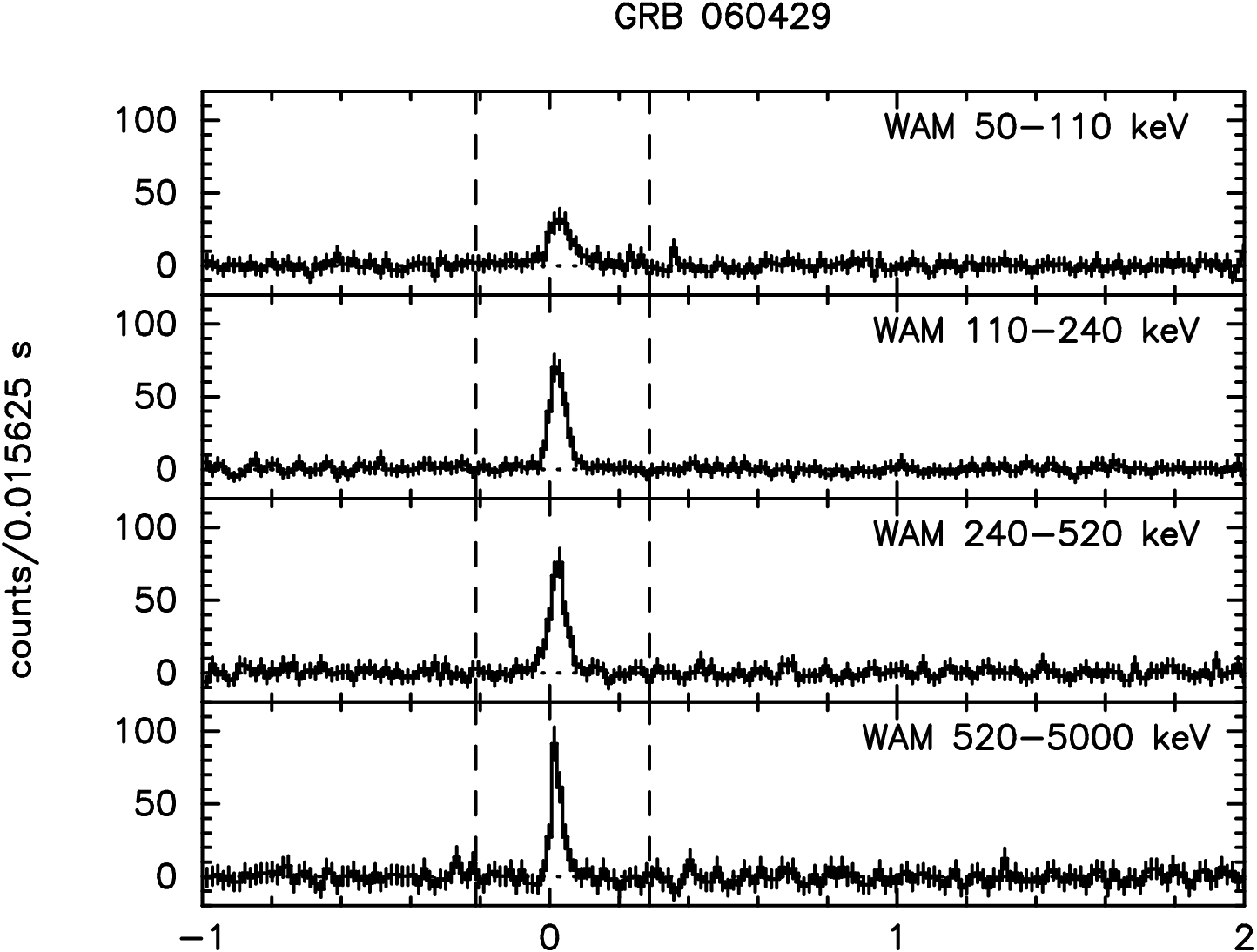}}}
\rotatebox{-90}{\resizebox{!}{8cm}{\includegraphics{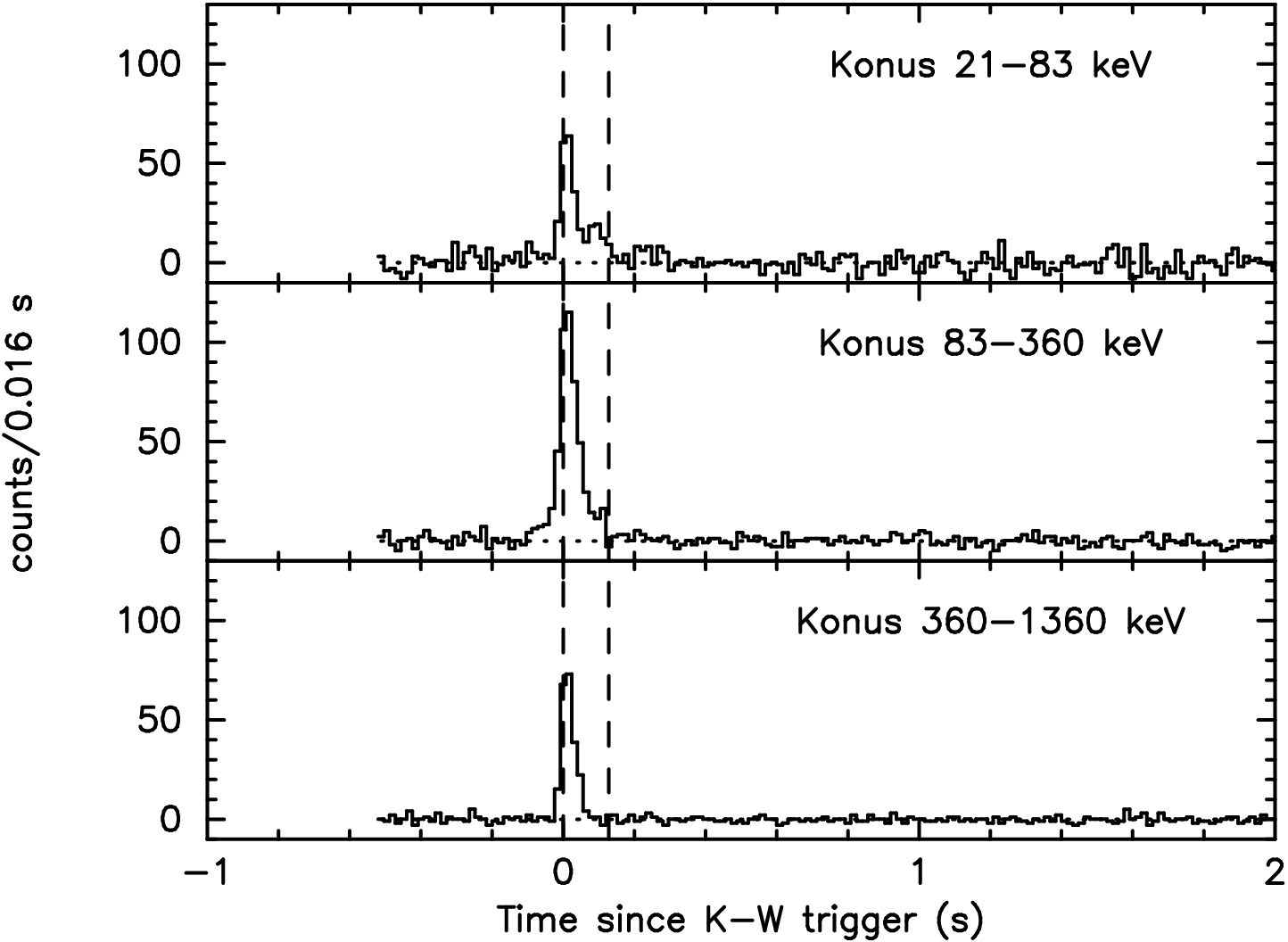}}}
\end{center}
\begin{center}
\rotatebox{-90}{\resizebox{!}{8cm}{\includegraphics{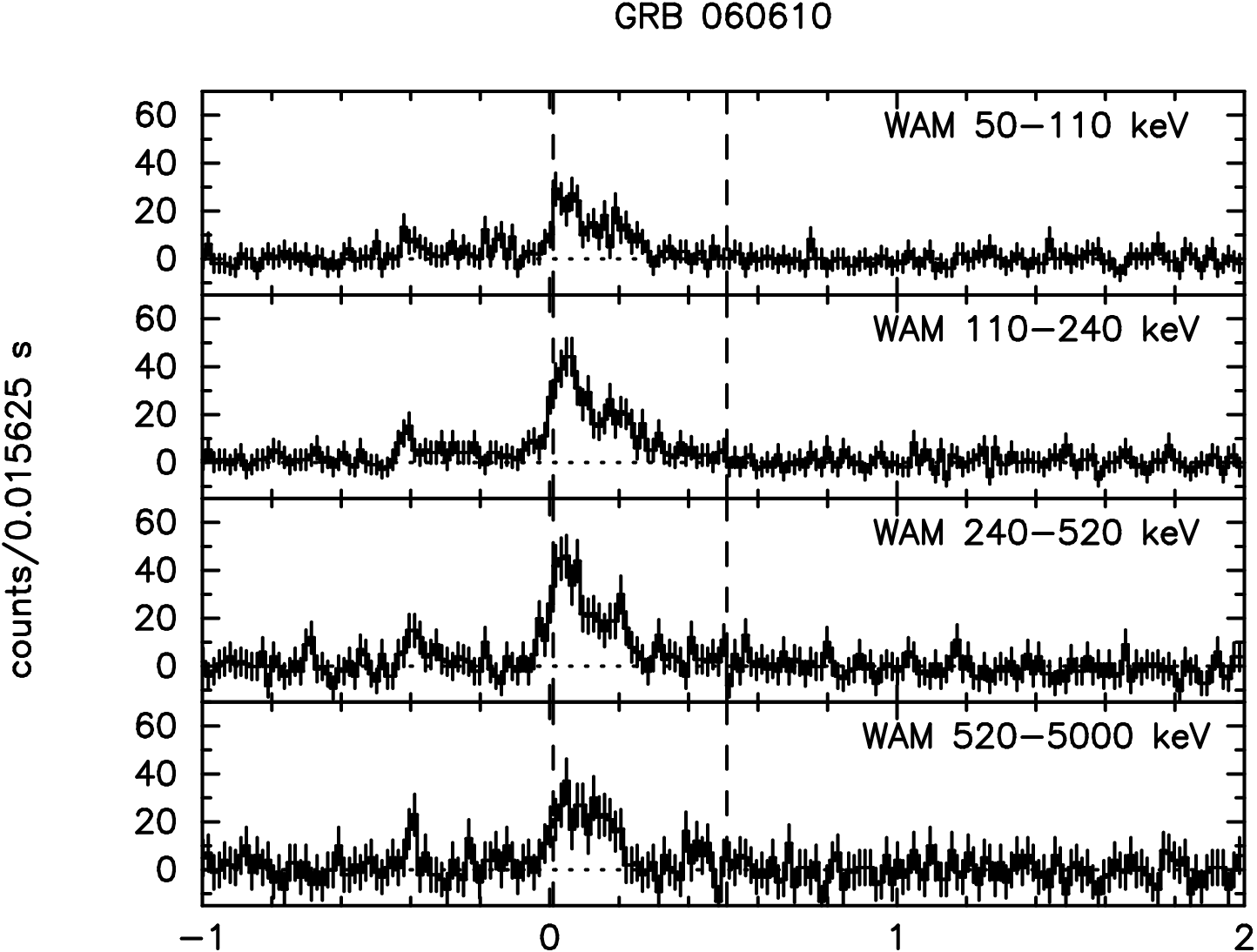}}}
\rotatebox{-90}{\resizebox{!}{8cm}{\includegraphics{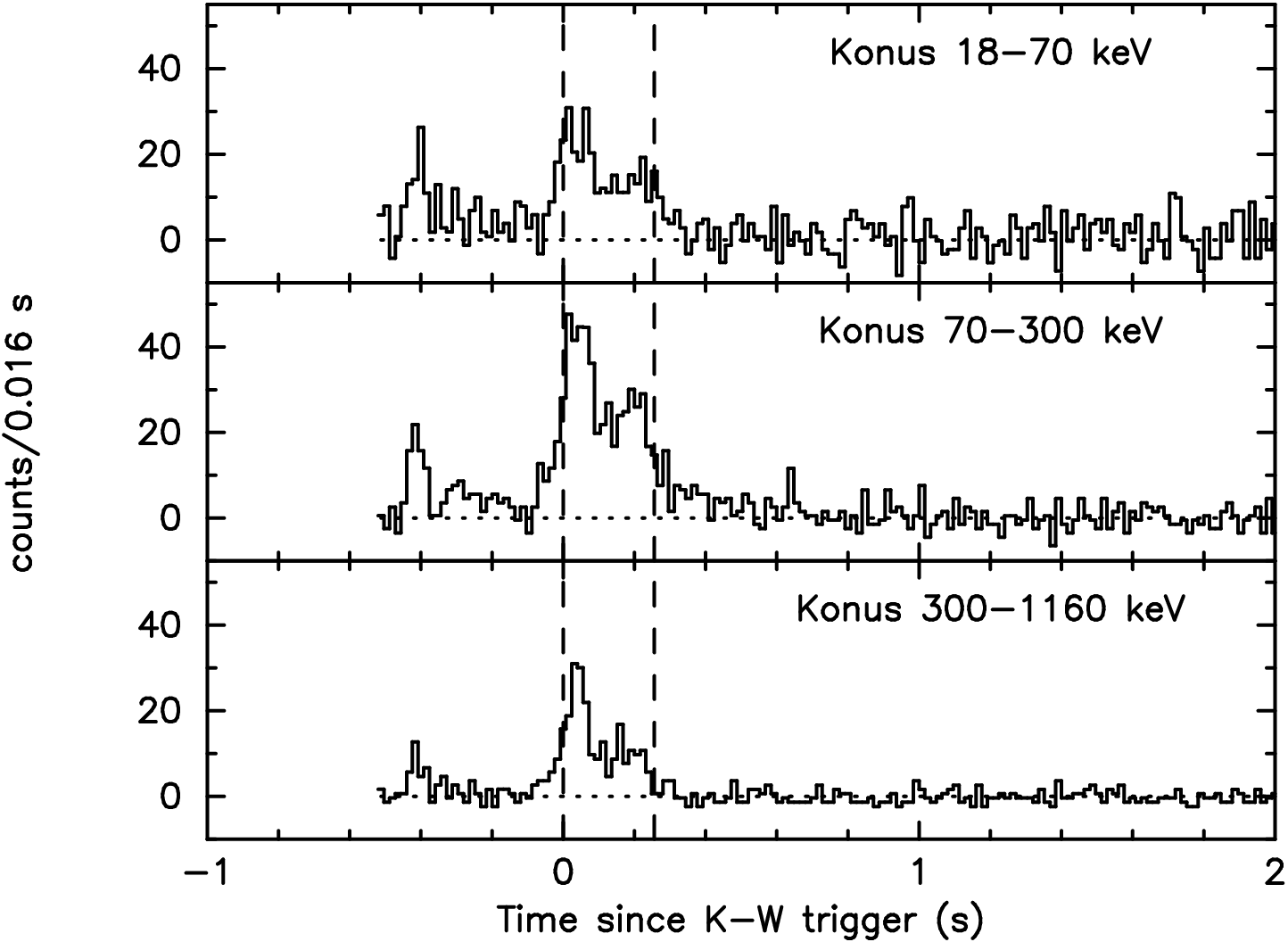}}}
\end{center}
\end{minipage}
\caption{Background-subtracted WAM (top) and Konus (bottom) light
 curves of our sample of short GRBs. The zero times of these light
 curves are adjusted to the Konus trigger time.
The time interval for extracting the spectrum is shown by the
 dashed-line. Left-top panel is GRB 051127, right-top panel is GRB
 060317, left-bottom panel is GRB 060429, and right-bottom panel is GRB
 060610.}
\label{lightcurve}
\end{figure}

\begin{figure}[htbp]
\begin{minipage}{8cm}
\begin{center}
\rotatebox{-90}{\resizebox{6cm}{!}{\includegraphics{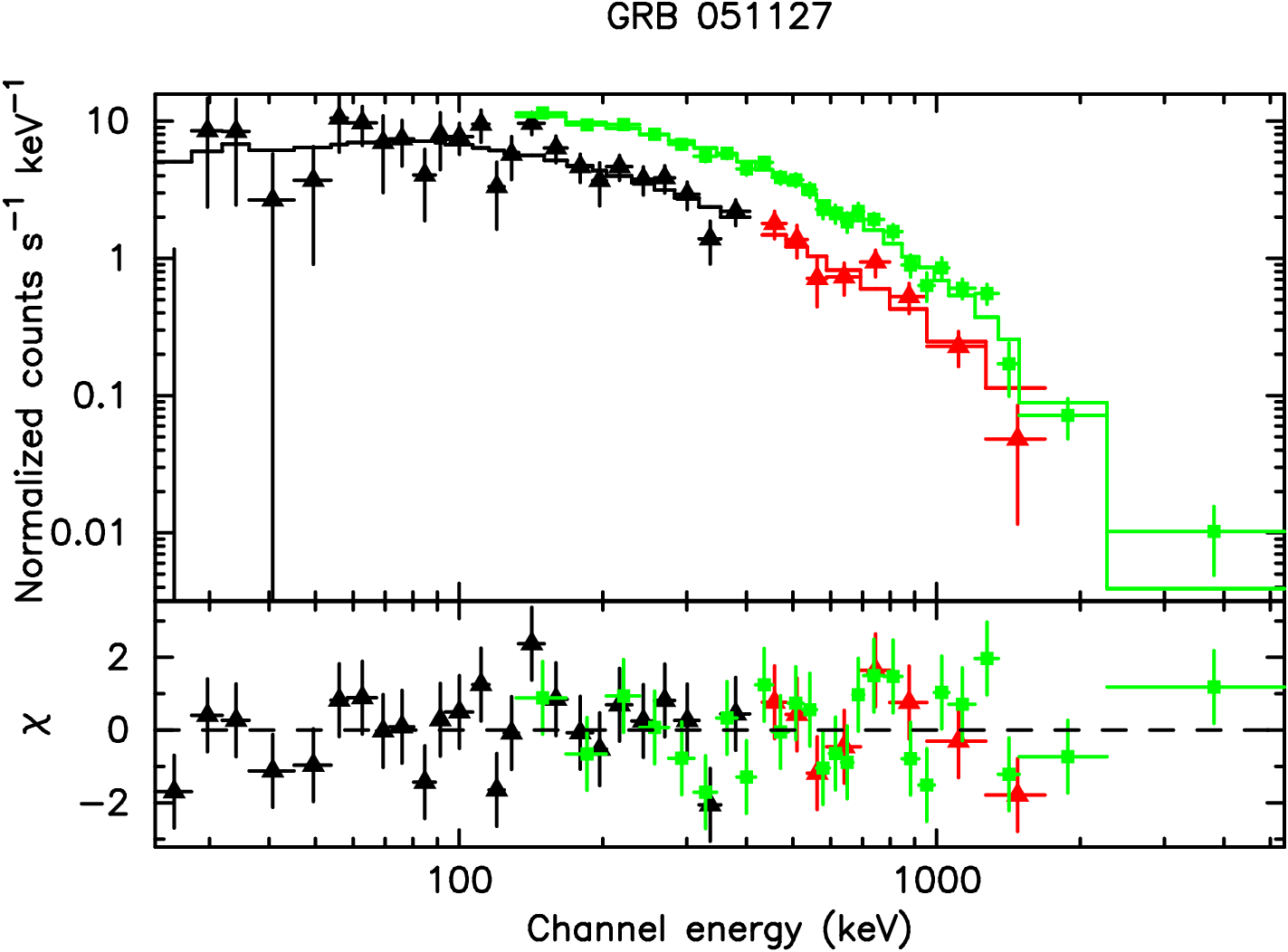}}}
\rotatebox{-90}{\resizebox{6cm}{!}{\includegraphics{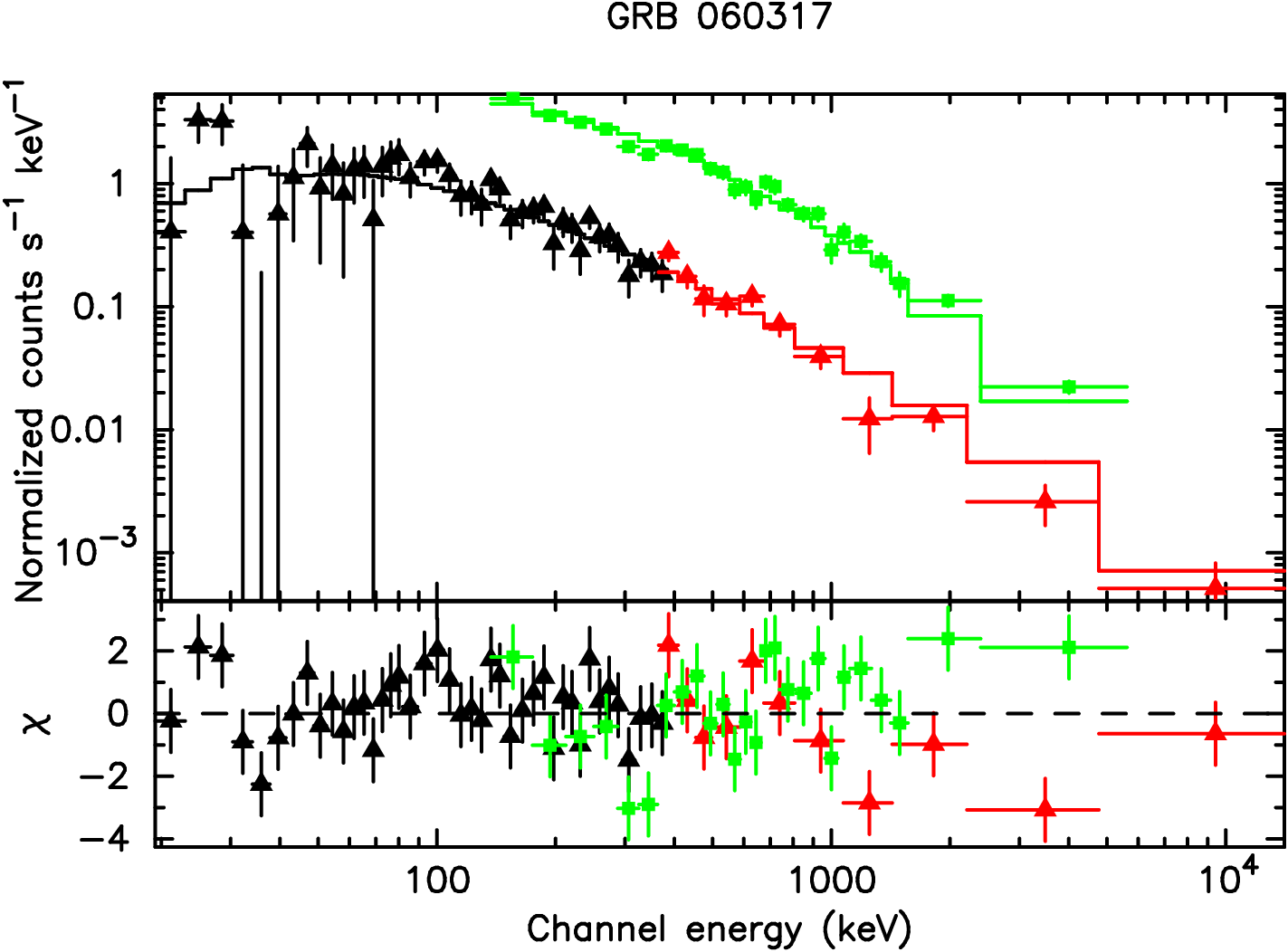}}}
\end{center}
\end{minipage}
\begin{minipage}{8cm}
\begin{center}
\rotatebox{-90}{\resizebox{6cm}{!}{\includegraphics{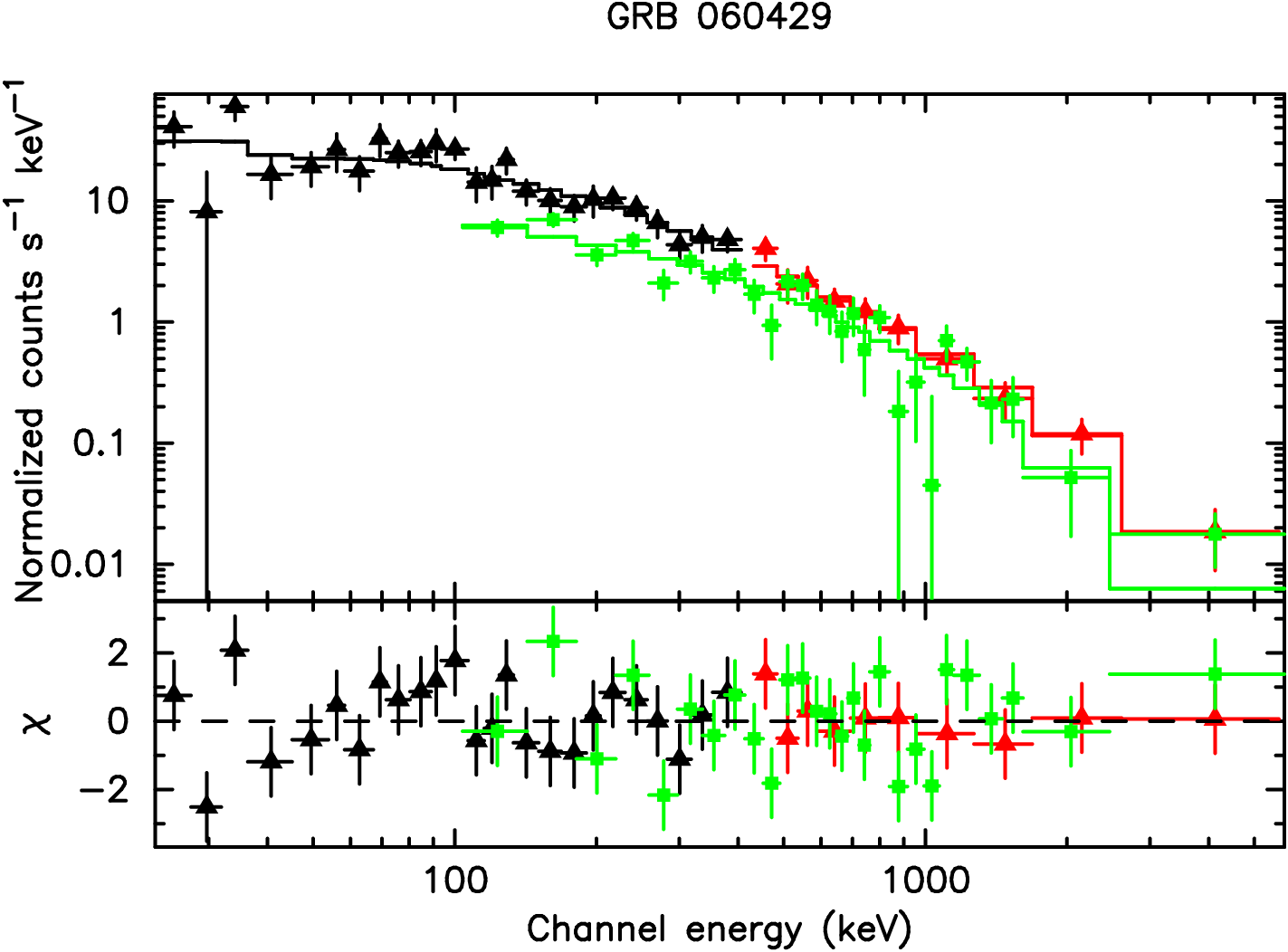}}}
\rotatebox{-90}{\resizebox{6cm}{!}{\includegraphics{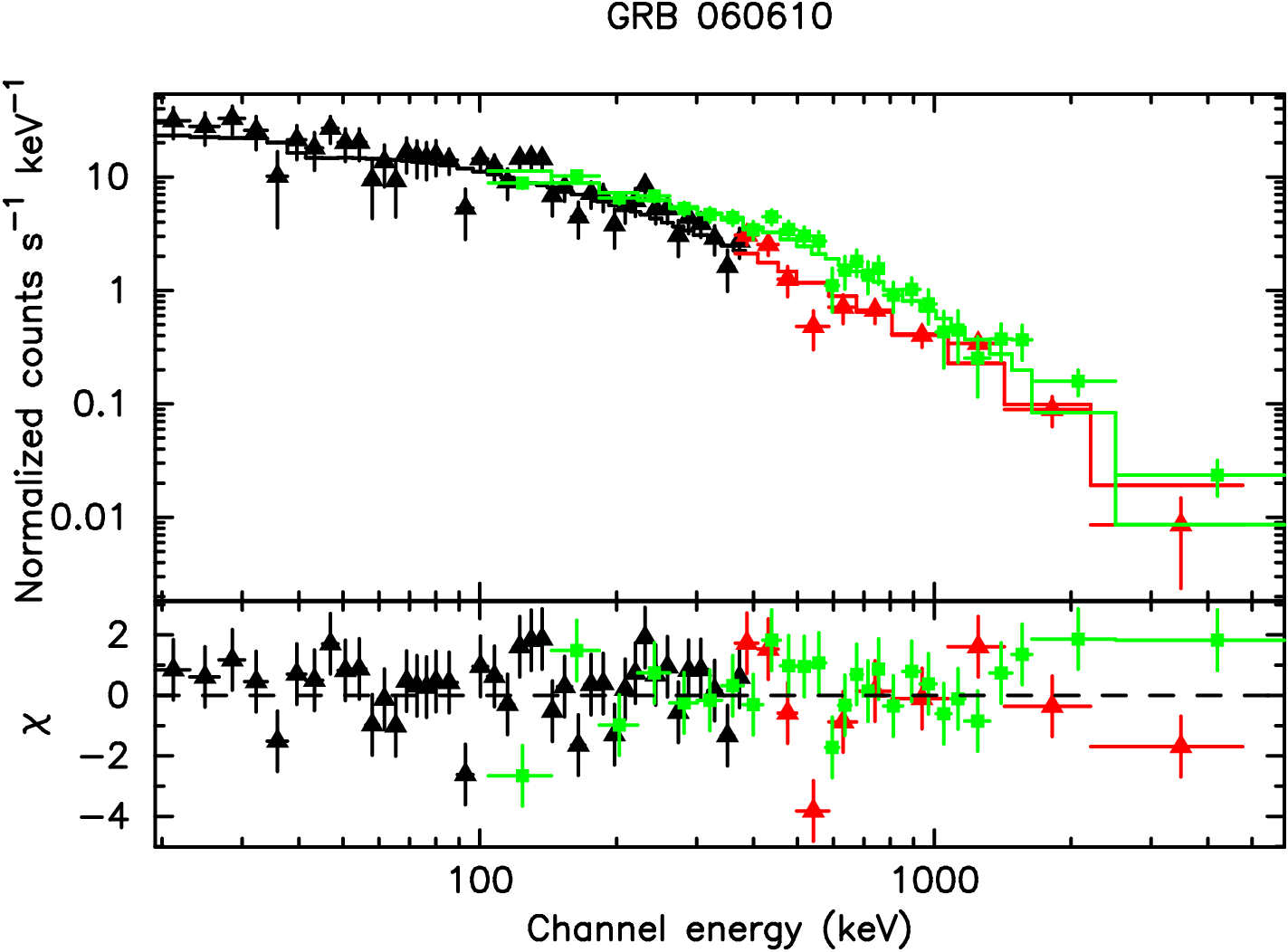}}}
\end{center}
\end{minipage}
\caption{Joint spectral fitting with a cut-off power law (CPL) model for the
 WAM (square) and the Konus (triangle) data. The solid lines represent
 the best-fit CPL model.}
\label{spectrum}
\end{figure}

The WAM and Konus light curves of GRB 051127 are shown in figure \ref{lightcurve}. 
The WAM light curve contains 
multiple spikes, and the T$_{90}$ duration (the time to accumulate
between 5 and 95\% of the counts), which is measured between 50
and 5000 keV, is 0.66 s. Thus, this burst 
is classified as short, using the burst 
duration alone. We extracted the spectrum from
T$_0$(WAM) to T$_0$(WAM)+1.0 s for WAM and T$_0$(KW) to T$_0$(KW)+0.256 s for
Konus, where the T$_0$s are summarized in table \ref{ipntable} and 
shown in figure \ref{lightcurve}.
The propagation delay from Suzaku to Wind is 4.364 s for this GRB,
i.e., correcting for this factor, one sees that the
T$_0$(KW) = T$_0$(WAM)+0.328 s.
We cannot adjust the time region further because the next Konus time
bin has a duration of 8.192 s.
The emission is clearly seen up 
to 5 MeV for the WAM and up to 2 MeV for Konus, as shown in figure \ref{spectrum}. 
We thus performed a fit from 100 keV to 5 MeV for the 
WAM and from 20 keV to 2 MeV for Konus. 
The WAM spectrum is well described by the CPL model with 
a photon index of $-$0.50$^{+0.18}_{-0.18}$ and a peak energy
E$_{\rm peak}$ of 1257$^{+280}_{-192}$ keV. For Konus, the same model
with a photon index of $-$0.29$^{+0.29}_{-0.21}$ and an E$_{\rm peak}$ of 940
$^{+280}_{-180}$ keV provides the best fit. The WAM spectrum seems to be
slightly softer than that of Konus. We believe that this is because
it includes a soft emission peak around T = T$_0$(WAM)$-$0.2 s. 
There are no significant differences in $\chi^2$ between the CPL and the Band
model fits for either spectrum.
The power-law photon index, $\alpha$, in the low-energy portion 
and the peak energy, E$_{\rm peak}$, are consistent 
with those obtained by the CPL model. However,
we cannot constrain the power-law photon index, $\beta$, at 
high energies 
and obtain only an upper limit of $\beta <$ $-$2.03. 
We then performed a joint fit. The CPL model with photon index 
$-$0.44$^{+0.15}_{-0.15}$ and E$_{\rm peak}$ 1168$^{+207}_{-150}$ keV
provides the best fit.  Table \ref{fit_summary_tbl} summarizes the
results of all of the conducted fits.

\subsection{GRB 060317}

There are two intense pulses in the light curve of GRB 060317, as seen in 
figure \ref{lightcurve}. The burst has a total duration of about
1.31 s for the WAM. 
We cannot distinguish whether this GRB belongs to the short or long
population from the burst duration alone, because the duration
distribution overlaps around 1--2 s. 
In order to extract the spectrum from both pulses, the WAM time region 
was taken to be 
T$_0$(WAM) to T$_0$(WAM)+4.0 s.
For Konus, we had to select a larger time region
than that of the WAM, so as to
include both pulses: 
from T$_0$(KW) to T$_0$(KW)+8.336 s. 
Konus detected this burst 3.49 s before the detection by the
WAM, and thus T$_0$(KW) corresponds to T$_0$(WAM)+0.389 s.
We can see that
emission is clearly detected up to 5 MeV for WAM and up to 10 MeV for
Konus, as shown in figure \ref{spectrum}.
Therefore, we performed the spectral fitting from 100 keV to 5
MeV for the WAM and from 20 keV to 10 MeV for Konus. 
We cannot constrain E$_{\rm peak}$ from the WAM spectrum, 
and a simple power-law with a
photon index of $-$1.11$^{+0.05}_{-0.05}$ fits
the observed spectrum well.
On the other hand, the Konus spectrum is well described by the
CPL model with a photon index of $-$1.02$^{+0.15}_{-0.13}$ and 
an E$_{\rm peak}$ of 2089$^{+1199}_{-634}$ keV. The Band model does not
provide any improvement to the fit. When we performed the joint
fit, E$_{\rm peak}$ became very large
(5687$^{+3032}_{-1672}$ keV). This might come from the large
IPN error box, which is about 5 deg$^2$, and the systematic 
uncertainty might be larger than our estimation. Therefore, we take the
spectral parameters of this burst to be those obtained by Konus.

\subsection{GRB 060429}

This GRB shows  a very simple short spike structure with a duration of 
0.08 s in the WAM light curve (figure \ref{lightcurve});
based on duration alone, it
is short enough to
be placed in the short class. 
Since we changed the WAM time resolution from 1.0 s to 0.5 s for 
spectral data, and from 1/32 s to 1/64 s for 
light curve data on 2006 March 20, 
we could select the time region to be 
T$_0$(WAM) to T$_0$(WAM)+0.5 s,
and obtain a better S/N. For 
Konus, we selected the time interval from 
T$_0$(KW) to T$_0$(KW)+0.128 s.
The propagation delay from Suzaku to Wind is 1.530 s for this GRB,
i.e., correcting for this factor, one sees that the
T$_0$(KW) = T$_0$(WAM)+0.211 s.
We can see in figure \ref{spectrum} that the emission extends up to 5 MeV for
both WAM and
Konus. We thus performed 
spectral fittings from 100 keV to 5 MeV for the WAM and from 20 keV to 5
MeV for Konus, respectively.
These spectra are also well described by the CPL model with
a high E$_{\rm peak}$. The photon index is $-$0.99$^{+0.26}_{-0.28}$ for
the WAM and $-$0.76$^{+0.15}_{-0.13}$ for Konus, respectively.
The E$_{\rm peak}$ can be constrained to 1583$^{+429}_{-371}$ keV 
by Konus, but the WAM data do not constrain it. The fit does not
improve in the Band model. The joint fit also
gives  consistent results.

\subsection{GRB 060610}

We can regard this GRB as short, based on the WAM burst duration of 0.79
s alone. 
The light curve exhibits a double-peak structure, as shown in figure
\ref{lightcurve}. The time regions for spectral analysis were
selected 
from T$_0$(WAM)+0.5 s to T$_0$(WAM)+1.0 s 
and from T$_0$(KW) to T$_0$(KW)+0.256 s. 
The propagation delay between these two instruments was 1.402 s,
i.e., correcting 
for this factor, one sees that the T$_0$(KW) = T$_0$(WAM)+0.490 s.
Figure \ref{spectrum} shows that the emission
extends up to 5 MeV for both instruments. We thus performed spectral
fittings from 100 keV (WAM) and from 20 keV (Konus) to 5 MeV. Both
spectra are also well described by the CPL model.
The WAM and Konus photon indices are $-$0.90$^{+0.17}_{-0.22}$
and $-$0.94$^{+0.12}_{-0.11}$, respectively.
The E$_{\rm peak}$ is constrained by the WAM and Konus to 
2889$^{+2973}_{-1175}$ keV and 1536$^{+500}_{-366}$ keV, respectively. From the
joint fit, 
we obtained a photon index of $-$0.92$^{+0.09}_{-0.12}$ and
an E$_{\rm peak}$ of 1821$^{+495}_{-364}$ keV. 
There is no significant improvement of the fit with the Band
model, and we can not constrain the
power-law photon index, $\beta$, at high energies;
we obtained an upper limit of $\beta <$
$-$2.50.

\subsection{Spectral Lag}

\begin{figure}[H]
\begin{minipage}{8cm}
\rotatebox{-90}{\resizebox{3cm}{8.2cm}{\includegraphics{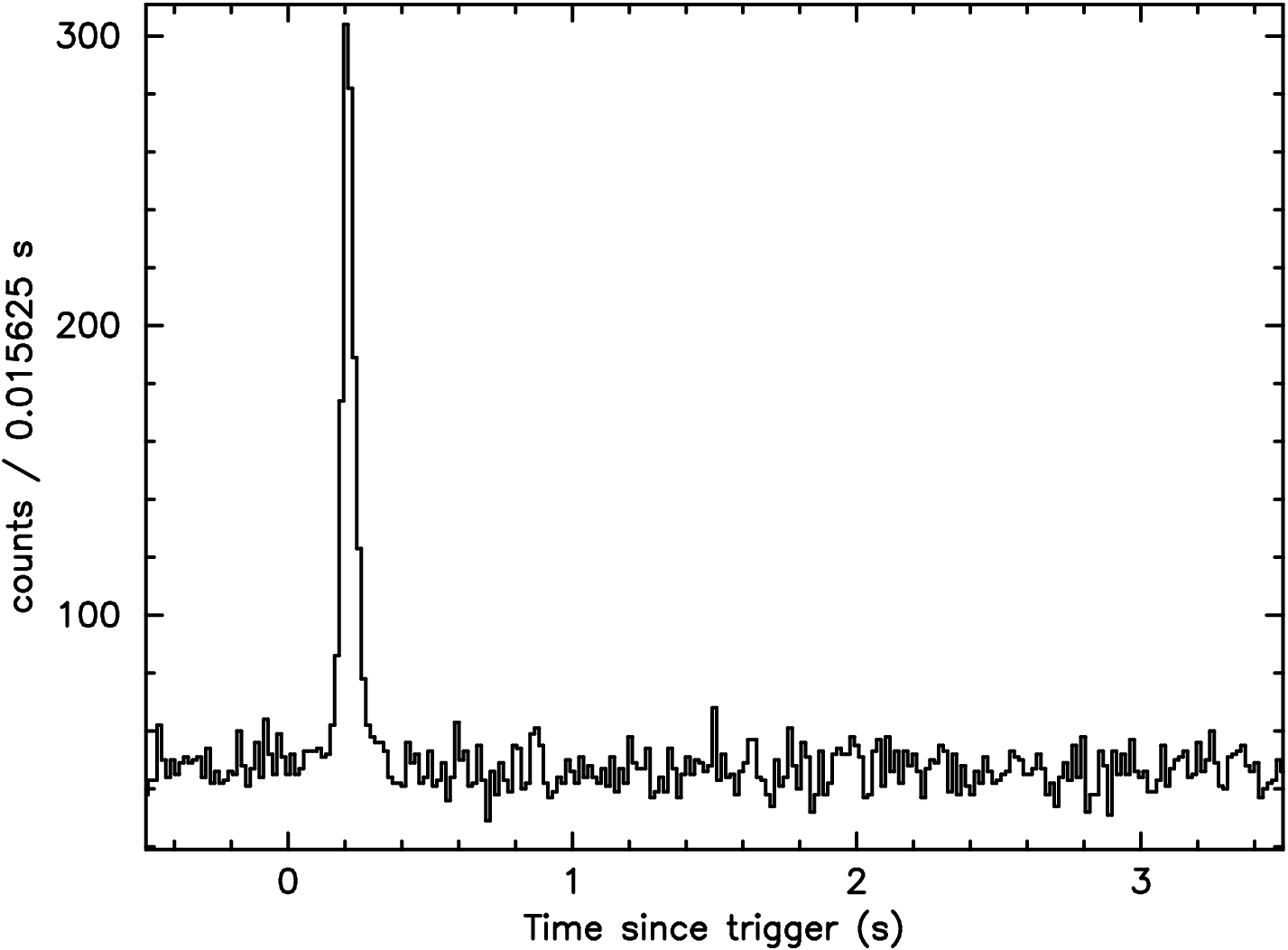}}}
\rotatebox{-90}{\resizebox{6cm}{!}{\includegraphics{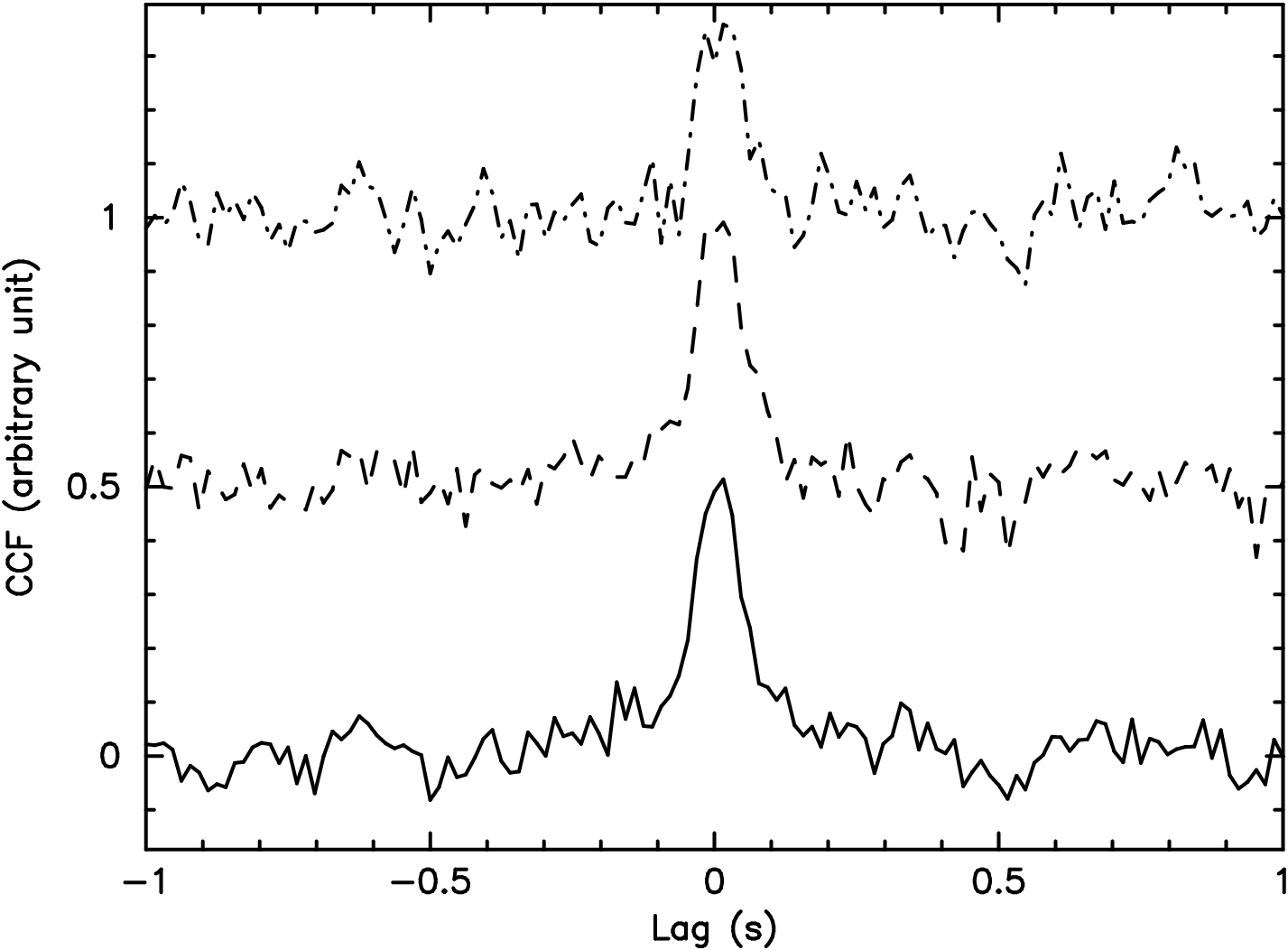}}}
\end{minipage}
\begin{minipage}{8cm}
\rotatebox{-90}{\resizebox{3cm}{8.2cm}{\includegraphics{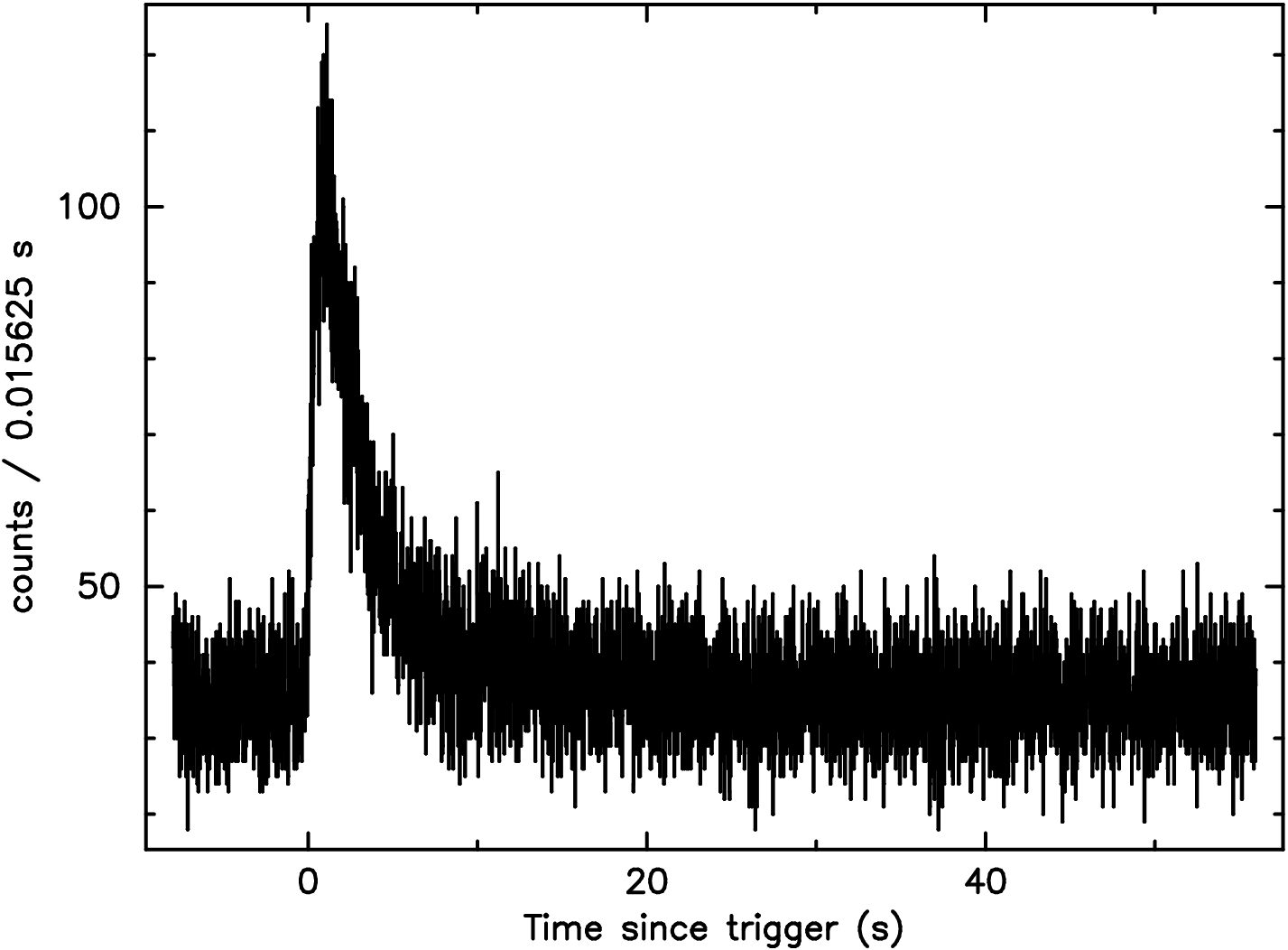}}}
\rotatebox{-90}{\resizebox{6cm}{!}{\includegraphics{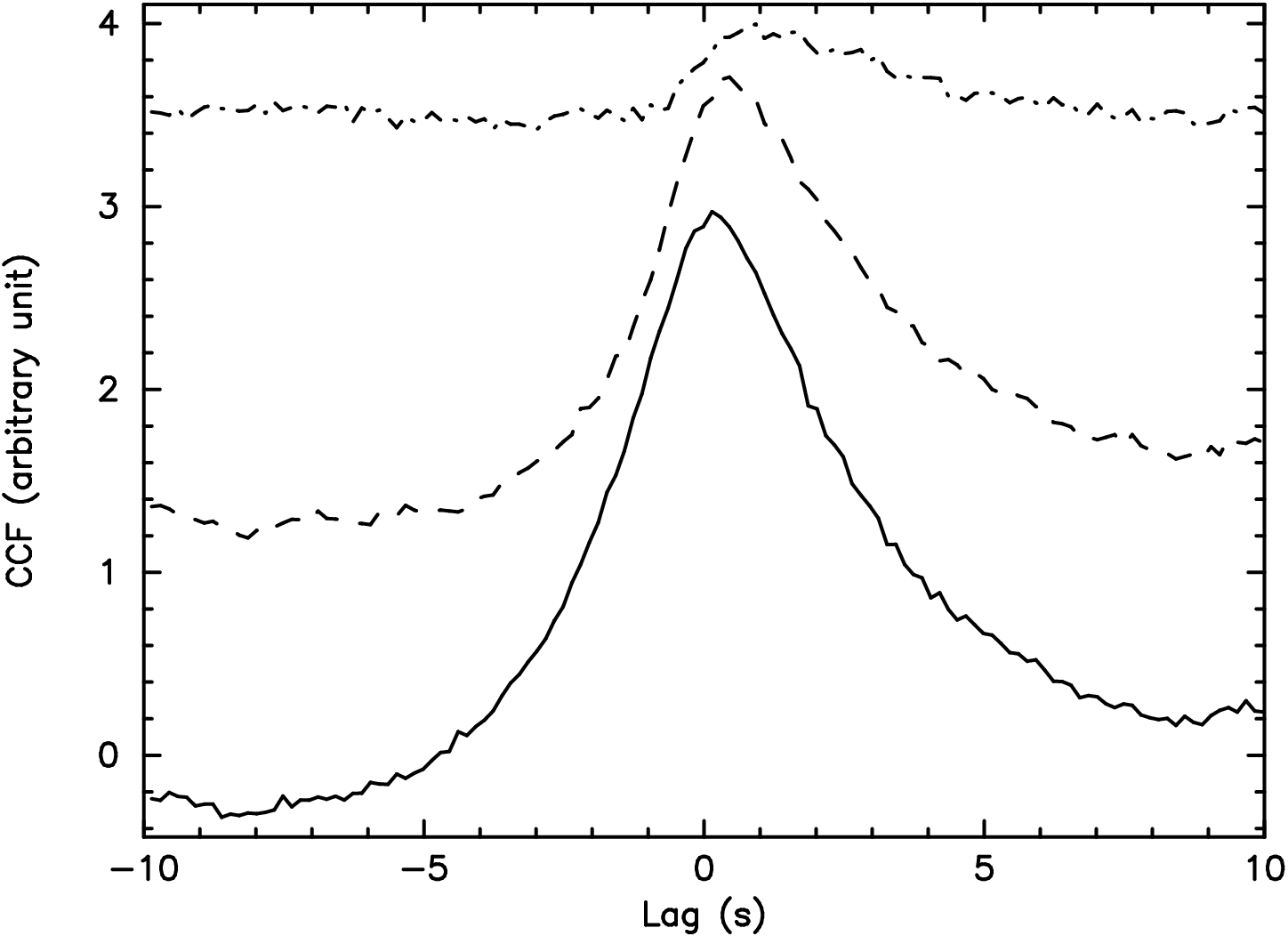}}}
\end{minipage}
\caption{Example of the light curves of the WAM data (top panel) and  the
 CCFs (bottom panel) for a short GRB 060429 (left) and a long GRB 060922 (right).
 The solid line shows CCF(50--110 keV to 110--240 keV), the dashed line shows
 CCF(50--110 keV to 240--520 keV), and the dot-dashed line shows CCF(50--110
 keV to 520--5000 keV). }
\label{ccf_lc}
\end{figure}

Figure \ref{ccf_lc} shows an example of the cross correlation function (CCF)
calculated for GRB 060429. The CCFs of our four short GRBs all exhibit
a sharp shape structure around zero. We estimate the peak values of the
CCFs using Gaussian profiles (\cite{Yi 2006}). The results are given in table \ref{temporal_summary_tbl}. The
spectral lags are all consistent with zero for all
energy ranges. For a comparison, we also calculated the CCF of three bright
long WAM GRBs:
050924, 060915, and 060922. Their CCFs display
broad peaks,  as shown in figure \ref{ccf_lc}. Thus, we used a polynomial instead of
a Gaussian profile to estimate the peak values; the results are also given in table \ref{temporal_summary_tbl}.
We found that these long GRBs have significant positive lags, 
which become larger towards the higher energy bands.
We confirmed that
this lag is not caused artificially, by analyzing
simulated data with no spectral lag, and indeed obtained zero lag.

\section{Discussion}

\subsection{Spectral Characteristics of Short GRBs}

The spectra of all four GRBs are well 
described by a power law with an exponential cut-off (CPL), and the fits do not improve significantly 
when we apply the Band model. 
To investigate the possibility that the spectra are better described by
the Band model, we used the Band model with a typical value of the power
law photon index in the high-energy portion of $\beta$ = $-$2.3. When we performed
 joint fits with this model, the $\chi^2$/d.o.f became 66/55, 
69/58, and 107/75 for GRB 051127, 060429, and 060610, 
respectively. For GRB 060317, there was also no improvement in the
Konus spectrum with this model, whose $\chi^2$/d.o.f is 53/49 (see table \ref{fit_summary_tbl}).
Therefore, we cannot determine whether these GRBs follow the CPL
model or the Band model from our data. 
The power-law photon index, $\alpha$, of the low-energy part of our sample
is distributed around $-$1.0 and the mean value is
$-$0.73. This is larger than that of the BATSE long duration GRBs of $-$1.05
(\cite{Ghirlanda 2002}).
In particular for GRB 051127, there is a significantly
flatter photon index of $-$0.41, which is larger than $-$2/3.
This distribution of $\alpha$ is also reported for BATSE bright,
short duration GRBs (\cite{Ghirlanda 2004}). They showed that the spectra of
their short GRB sample were also 
described by a CPL model.
If a flat slope is a  global
characteristic of the prompt emission spectrum of short GRBs,
 the simple synchrotron emission model may no longer be applied and 
other emission mechanisms, such as jitter radiation (\cite{Medvedev 2000}), 
are required.

\subsection{Classification of Short GRBs}

It is ambiguous in some cases to classify GRBs using the burst 
duration alone, because the observed long and short duration distributions 
overlap around 1.0--2.0 s. The
 dilation of the
intrinsic duration of GRBs due to cosmological distance, and the
dependence of duration on energy, must also be
considered.
Therefore, we utilize several other methods of classifying short and long
GRBs below. In this section, we discuss whether
the four bursts are really in the short class 
by examining these criteria: 1) spectral lag, 2) spectral hardness, 3)
radiation energy, and 4) extended emission.  
So far, these criteria have
been confirmed mainly based on the BATSE data. We can test these criteria in
a higher energy band than BATSE by the WAM and Konus data with high
statistics even in MeV energy band.

\begin{figure}[H]
\begin{center}
\rotatebox{-90}{\resizebox{7cm}{!}{\includegraphics{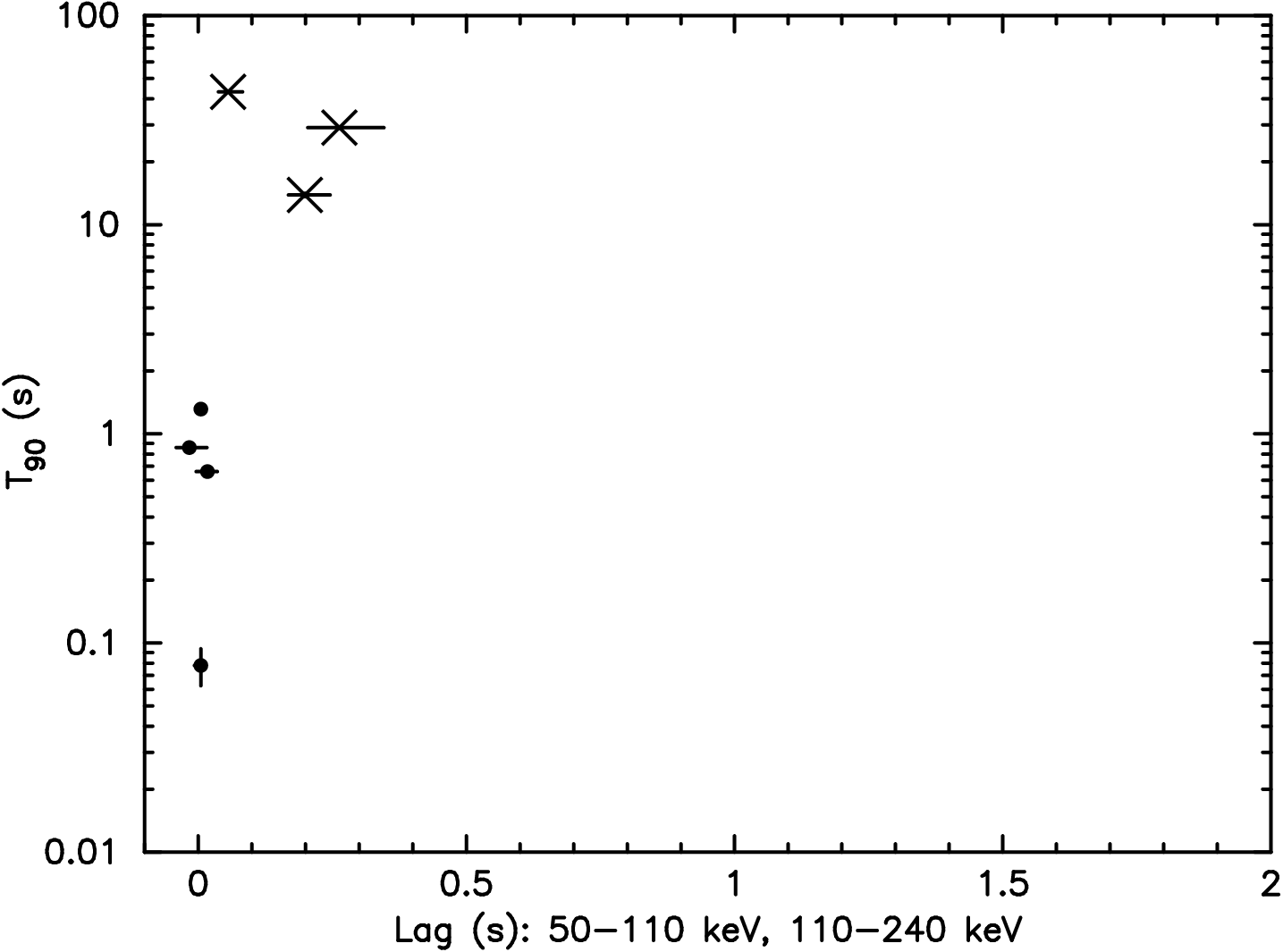}}}
\rotatebox{-90}{\resizebox{7cm}{!}{\includegraphics{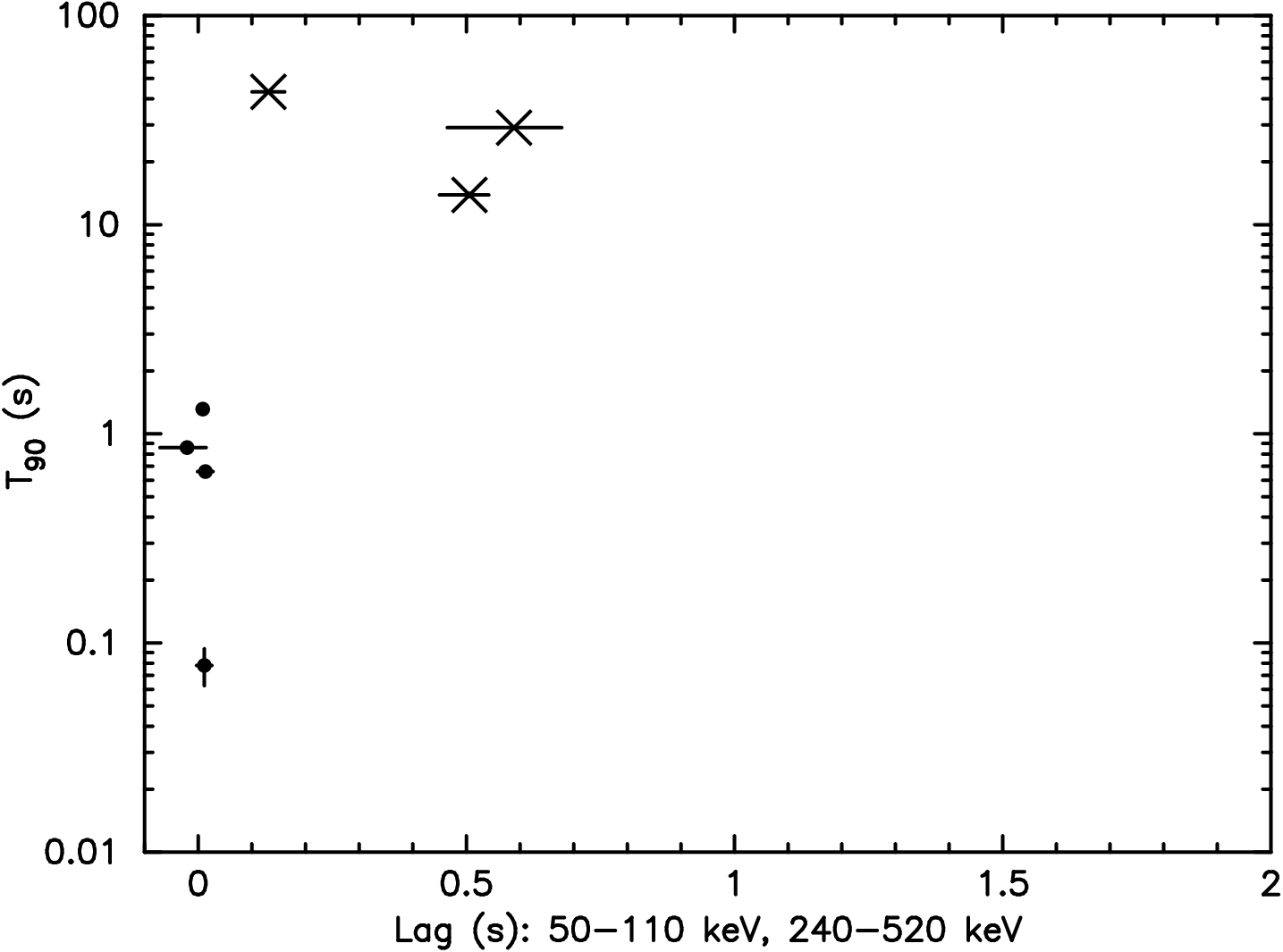}}}
\rotatebox{-90}{\resizebox{7cm}{!}{\includegraphics{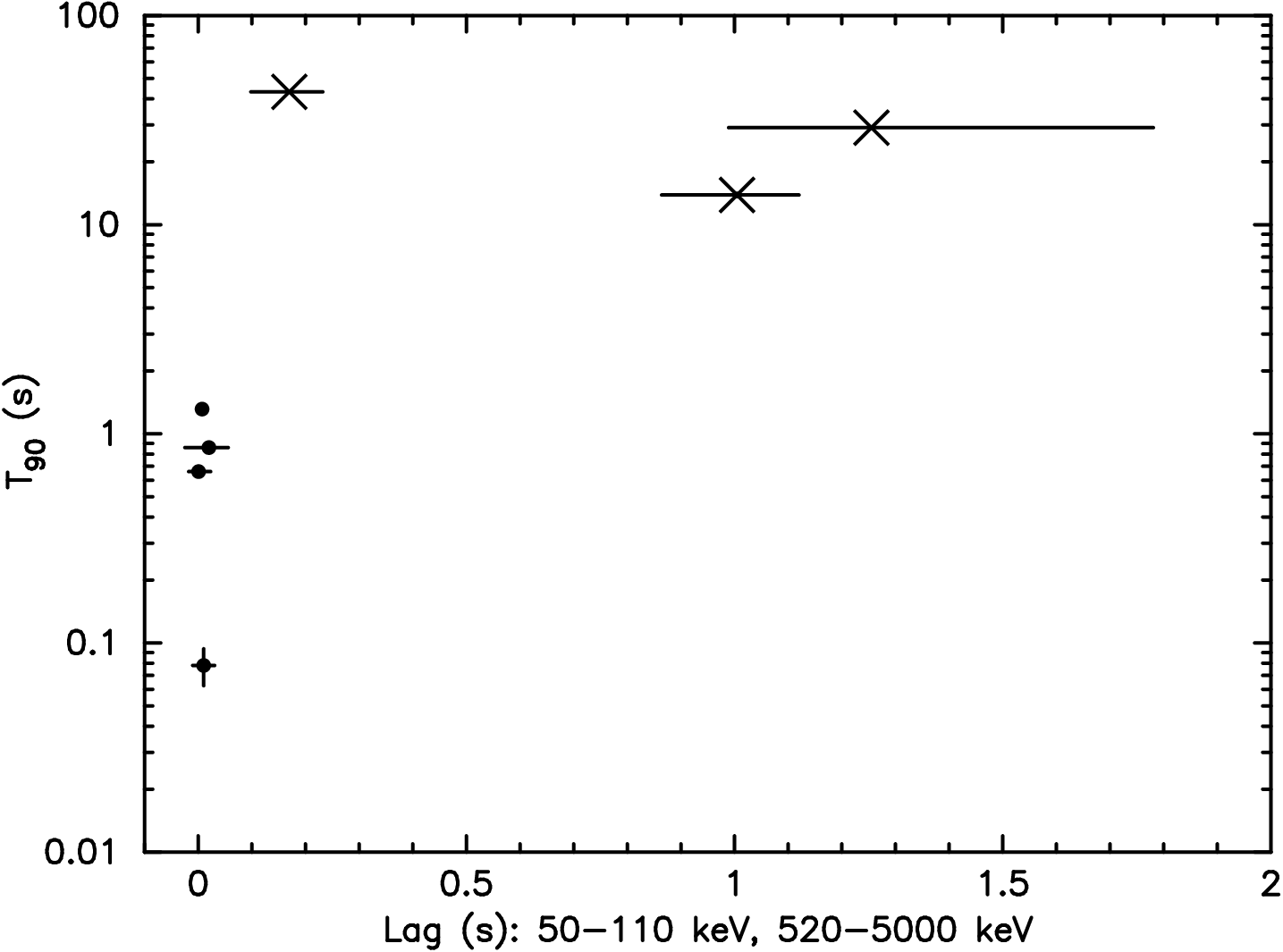}}}
\end{center}
\caption{Relation of the T$_{\rm 90}$ duration and the spectral
 lag. From top to bottom, the lag of 50--110 keV to 110--240
 keV, 50--110 keV to 240--520 keV, and 50--110 keV to 520--5000
 keV are shown. The filled circles are for the
 four short GRBs of this paper and the crosses are for long GRBs.}
\label{lag_vs_t90}
\end{figure}

The properties of the spectral lag are reported to differ 
between short and long-duration GRBs. 
Short GRBs exhibit lags, which are consistent with zero, 
    while long GRBs have significantly non-zero positive
     spectral lag (\cite{Norris 2000}; \cite{Norris 2006}). 
We have measured the spectral lags using the WAM light   curves.  
     Figure \ref{lag_vs_t90} plots the relation between the duration and
     the spectral lag for four short GRBs, for 50--110 keV vs. 110--240 keV,
     50--110 keV vs. 240--520 keV, and 50--110 keV vs. 520--5000 
     keV. We also plot the lags of three bright long duration
     WAM GRBs for comparison. 
     We find that the spectral lag of all four short
     GRBs is consistent with zero in approximately the same
     energy bands as BATSE 
    (50--110 keV vs. 110--240 keV or 50--110 keV vs.
     240--520 keV), while non-zero lag, evolving with $\rm T_{90}$, is seen for the three long
     GRBs. Therefore, the WAM data confirm the difference in the spectral
     lag obtained with BATSE. In addition, figure \ref{lag_vs_t90} 
demonstrates  the WAM capabilities in the high-energy band: lag evolution
     is seen above 520 keV for long GRBs.

\begin{figure}[H]
\begin{center}
\rotatebox{-90}{\resizebox{8cm}{!}{\includegraphics{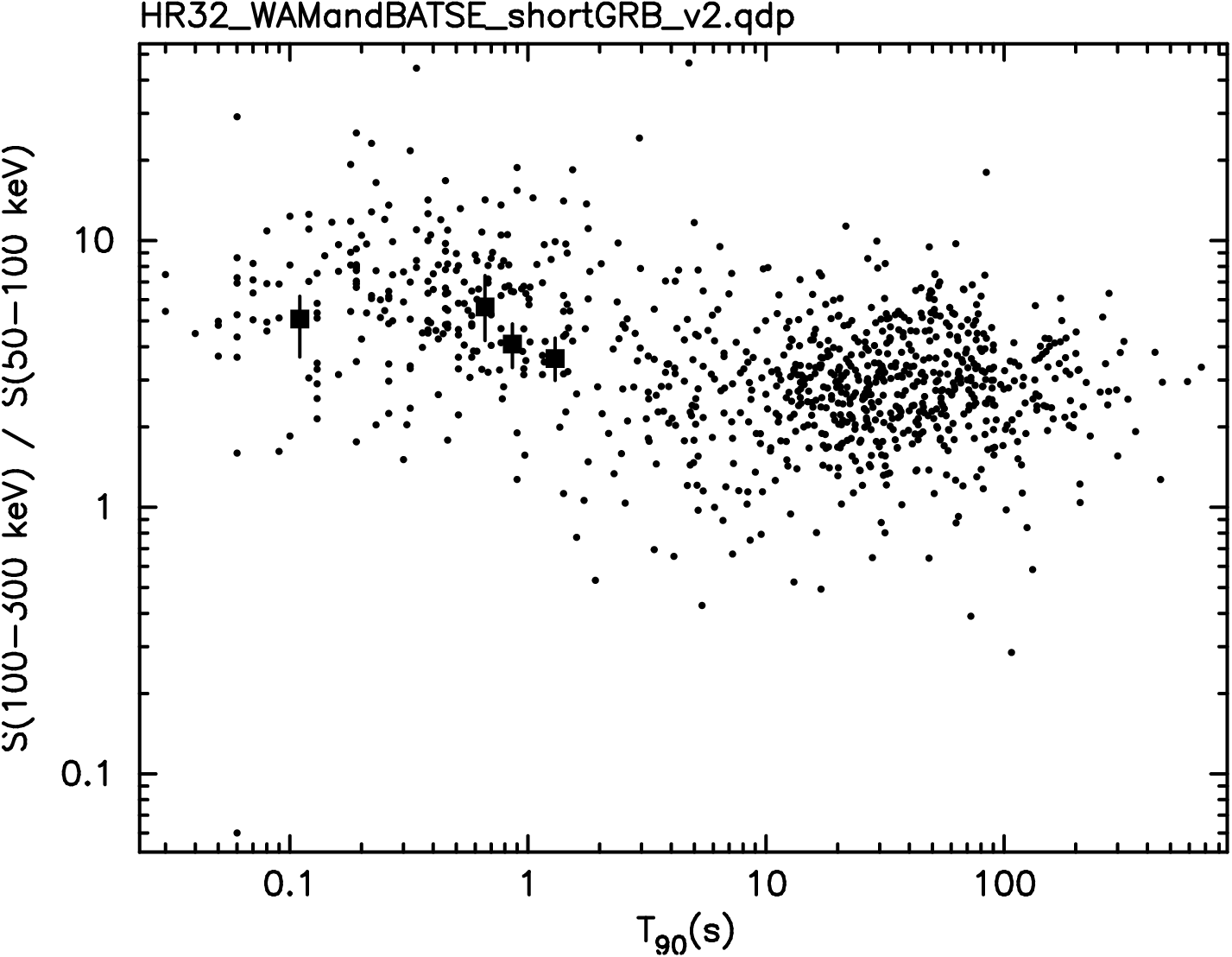}}}
\end{center}
\caption{Relation between the duration (T$_{90}$) and the hardness ratio
 of 100--300keV to 50--100 keV. The Dots show BATSE results. The results of our four short GRBs are shown as filled squares.}
\label{t90_vs_hardness}
\end{figure}

The hardness ratio of short GRBs is reported to be slightly
harder than that of long GRBs from BATSE results (\cite{Cline 1999};
\cite{Paciesas 1999}). 
Although the boundary between short and long GRBs 
in the duration-hardness plane is ambiguous, we nevertheless
     consider this difference as a possible criterion. 
Therefore, we obtained the spectral hardness of our sample of short
     GRBs, based on the best-fit spectral model. We derived the
     hardness ratio using the 50--100 and 
100--300 keV fluence, which is same energy range as BATSE.     
We plot them against the  T$_{90}$ duration in figure 
 \ref{t90_vs_hardness}, together with the BATSE results
     (\cite{Paciesas 1999})
 The hardness ratios of our
     short GRBs are distributed from 3.61 to 5.64 with an average value of 4.61,
     and all have greater hardness ratios
     than the average value of the BATSE long (T$_{90} >$ 2.0) GRBs (3.27). 
Therefore, all four are consistent with a short classification.
    However, these short GRBs do not show a clear separation from the long
     GRBs, as in the BATSE samples in figure
     \ref{t90_vs_hardness}.
     Although all our short GRBs have an E$_{\rm peak}$ around 1 MeV,
     none of them exceeds a hardness ratio of 8, whereas 
     half of the BATSE short GRBs do. This is because the hardness
     ratio depends on the low-energy photon index rather than E$_{\rm peak}$ 
     (\cite{Ghirlanda 2004}; \cite{Sakamoto 2006}). The short GRB which
     has the highest hardness ratio in our sample is GRB 051127, with 5.64
     $^{+1.75}_{-1.42}$. GRB 051127 also has the hardest photon index, $-$0.44, but
     its E$_{\rm peak}$ is one of the lowest in our sample. As mentioned in
     \cite{Sakamoto 2006}, the low-energy photon index has to be flatter
     than 0 to reproduce a hardness ratio greater than 8, which is seen in the
     BATSE ``very hard'' short GRBs.
     Based on our broad band spectroscopy of bright, high E$_{\rm
     peak}$ short GRBs, none has a hardness ratio similar
     to the BATSE ``very hard'' short GRBs. 

\begin{figure}[htbp]
\begin{center}
\rotatebox{-90}{\resizebox{8cm}{!}{\includegraphics{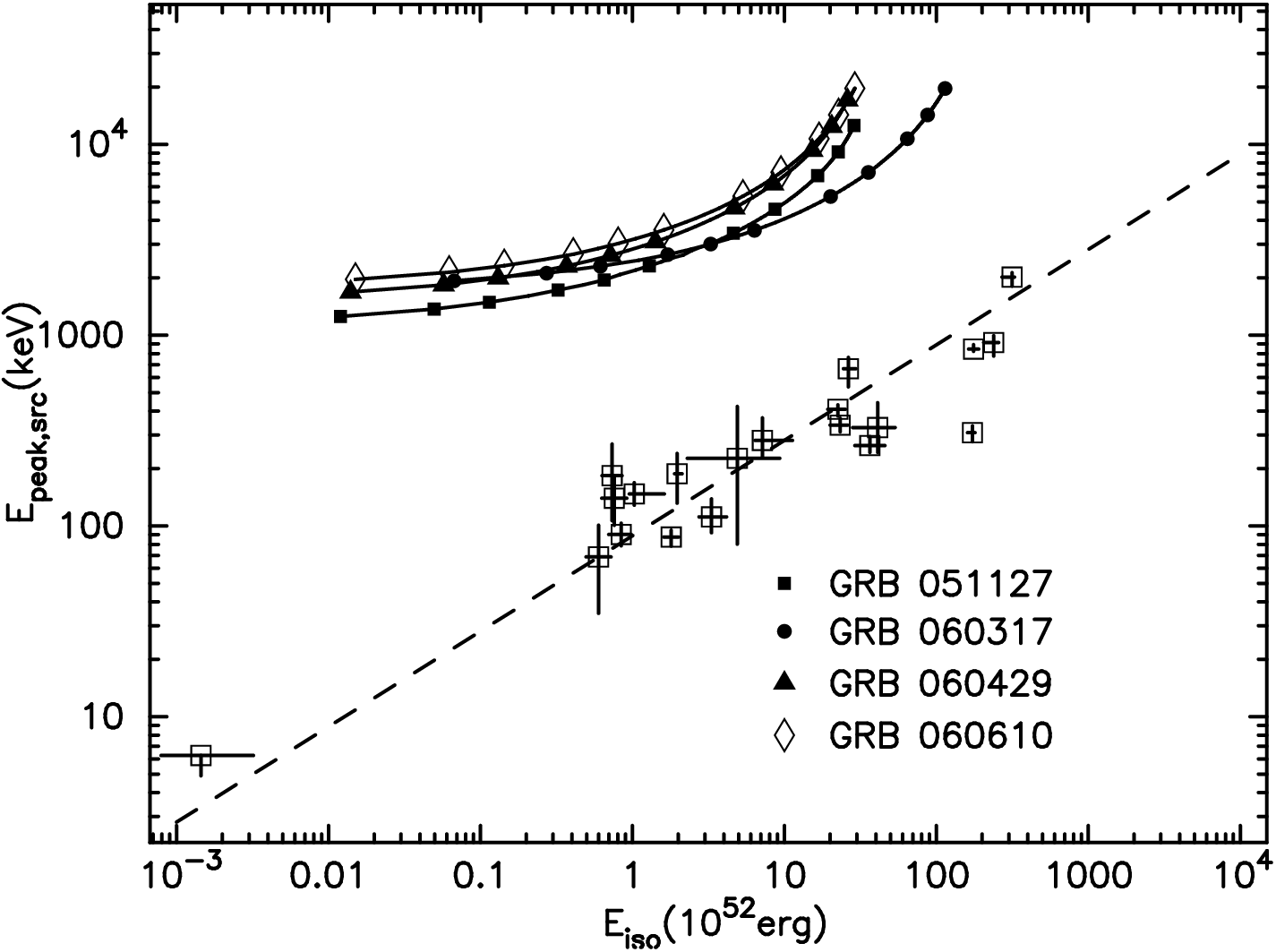}}}
\end{center}
\caption{Relation between E$_{\rm peak}$ and E$_{\rm iso}$ for
our four short GRBs (filled square, GRB 051127;
 filled circle, GRB 060317; filled triangle, GRB 060429; and open
 diamond, GRB 060610). The trajectories were calculated by assuming redshifts 
from 0.1 to 10.0. The data points along the trajectories correspond to
 z=0.1, 0.2, 0.3, 0.5, 0.7, 1.0, 2.0, 3.0, 5.0, 7.0, and 10.0,
 respectively. The dashed-line represents the Amati relation E$_{\rm
 peak, src}$ $\propto$ E$^{0.5}_{\rm iso}$ for long GRBs.}
\label{amati_relation}
\end{figure}

     For long GRBs, a correlation between the peak energy and the isotropic
     equivalent radiation energy in the source frame has been reported: 
     the Amati relation (\cite{Amati 2002}). 
     However, short GRBs whose redshifts are known have lower isotropic energy 
     than long GRBs and do 
     not satisfy this relation  (\cite{Amati 2006}). 
We calculated the isotropic energy, E$_{\rm iso}$,
     of our four short GRBs from 1.0 to 10000 keV in the source frame,
     assuming redshifts from 0.1 to 10.0. We used cosmological
     parameters (H$_0$, $\Omega_{\Lambda}$, $\Omega_m$) = (65, 0.7,
     0.3).  We included the effects of redshift on the energy 
    spectra and E$_{\rm peak}$ of these bursts.
     Figure \ref{amati_relation} shows the results of the calculation in
     the E$_{\rm p}$--E$_{\rm iso}$ plane together with the HETE-2 and
     BeppoSAX results (\cite{Sakamoto 2005b}). 
     We could not find any redshift that
     satisfies the 
    Amati relation. 
     All of the bursts in our sample exhibit a lower isotropic radiation energy by more
     than one order of magnitude when compared with long bursts of the same peak energy. 
     This is consistent with previous results.

     Long, soft emission humps a few tens of seconds after the short
     main spike were reported for several short GRBs (\cite{Norris 2006}),
     particularly GRB 050709 (\cite{Villasenor 2005}) and 050724
     (\cite{Barthelmy 2005a}), which have measurements of  the
     redshift and detections of the host galaxy. Thus the
     existence of the soft emission hump is considered to be one of the
     criteria to classify short GRBs. We searched for
     long soft humps in the time histories of our four short GRBs, but did not detect them. 
To do this,
     we calculated the upper limits to the peak soft photon flux at the 3
     $\sigma$ level for the WAM and Konus. In this calculation, we
     assumed that the spectrum of the soft emission was a simple power
     law with photon index $-$2.0. We found  3 $\sigma$
     upper limits to the 2.0 to 25 keV peak flux of
     4.47, 6.63, 9.21, and 7.54 photons s$^{-1}$ cm$^{-2}$, for GRB
     051127, 060317, 060429, and 060610, respectively,  
     using both the WAM and Konus.  
     All of these values exceed 2.72 ($\pm$ 0.47), which is 
      obtained for GRB 050709
     (\cite{Villasenor 2005}).
Therefore, the non-detection of the soft emission for these
bursts adds no constraint: 
the extended emission could be too soft to be detected by the WAM and
Konus.

\section{Conclusions}

We performed a joint analysis of four
bright, hard-spectrum, short-duration GRBs, localized by the IPN, with the Suzaku-WAM and 
Konus-Wind, in order to investigate the spectral
properties of short GRBs into the MeV energy band.
From the spectral analysis, we found that these bursts 
have a high E$_{\rm peak}$ around 1 MeV; 
the spectral parameters can be constrained tightly
by joint fitting. However, we could not determine whether these spectra
follow the CPL model or the Band model, because there 
is no  significant improvement between these models. 
The power-law photon index is slightly flatter than
that of long BATSE GRBs. In particular, 
GRB 051127 has a very flat spectrum with a photon index, $\alpha$, that is larger than
the synchrotron limit of $-$2/3.

We also examined several other spectral and temporal properties 
to confirm that the bursts belong to the short class.
Most properties, such as the spectral lag, the spectral hardness, 
and the isotropic radiation energy, indicated that the four bursts
were indeed in this class.
In addition, a broad-band spectral analysis with the WAM and Konus revealed
some interesting properties of short GRBs: i) There is no evidence for
very hard GRBs that have the large hardness ratios seen in BATSE
short GRBs. ii) The spectral lag of our short GRBs 
does not show any clear evolution, even in the MeV energy region, unlike that of
long hard GRBs. Although our short GRB sample is still limited, these
results might suggest one possibility, that the hardness ratios of the
short GRBs are not so different from those of long GRBs, and also
that the spectral lag at high energies can be a strong criterion for
classifying short and long GRBs.

\bigskip
M. O. is supported by the Research Fellowships of
the Japan Society for the Promotion of Science for Young Scientists (2006).
K. H. is grateful for IPN support under the NASA LTSA program, grant
FDNAG5-11451, the INTEGRAL Guest Investigator program, NAG5-12706 and
NNG06GE69G, and the Suzaku Guest Investigator program NNX06AI36G.
The Konus-Wind experiment is supported by
a Russian Space Agency contract and RFBR grant 06-02-16070. 
We are grateful to Richard Starr for providing the Mars Odyssey
data, and to Giselher Lichti, Arne Rau, and Andreas von Kienlin
for providing the INTEGRAL SPI-ACS data.
We would like to thank Jay Norris for useful advice concerning the
spectral lag analysis.

\begin{table}[htbp]
\caption{Interplanetary Network Spacecraft Observations of the four WAM and Konus
 short GRBs.}
\label{ipntable}
\begin{center}
\begin{tabularx}{17cm}{ccc}
\hline
\multicolumn{2}{c}{ Trigger time (UT)}       & Spacecraft$^{\ast,\dag}$ \\
Suzaku-WAM & Konus-Wind & \\
\hline
\multicolumn{3}{l}{GRB 051127}\\
22:55:19.896 & 22:55:15:860 & Konus, Suzaku, Odyssey HEND, HETE-FREGATE
 \\
\hline
\multicolumn{3}{l}{GRB 060317}\\
11:17:39.104 & 11:17:35.996 & Konus, Suzaku, INTEGRAL SPI-ACS \\
\hline
\multicolumn{3}{l}{GRB 060429}\\
12:19:51.031 & 12:19:49.712 & Konus, Suzaku, Odyssey HEND \& GRS,
 INTEGRAL SPI-ACS, RHESSI \\
\hline
\multicolumn{3}{l}{GRB 060610}\\
11:22:23.544 & 11:22:22.632 & Konus, Suzaku, Odyssey HEND \& GRS, INTEGRAL SPI-ACS, RHESSI \\
\hline 
\end{tabularx}
\end{center}
\scriptsize{$\ast$: HEND and GRS are the High Energy Neutron Detector and Gamma-ray Sensor
head (\cite{Hurley 2006a})}\\
\scriptsize{$\dag$: SPI-ACS is the spectrometer anti-coincidence detector (\cite{Rau 2005})}
\end{table}

\begin{table}[htbp]
\caption{Error box centers and corners of IPN localization and Incident angle.}
\label{ipntable2}
\begin{center}
\begin{tabularx}{15cm}{ccccccc}
\hline
GRB & Area, sq. deg. & $\alpha_{2000.0}$ & $\delta_{2000.0}$ &
 &\multicolumn{2}{c}{Incident angle [degree]$^\ast$}\\
&   &                &                 &                 &$\theta$, $\phi$
 (WAM)&$\theta$ (Konus)\\
\hline
051127 & 0.13& 301.500  & $-$37.782 & (center)  & 56.3, 80.2 & 17.4\\
       &     & 301.371  & $-$36.873 &           & 56.6, 79.1 &     \\
       &     & 301.695  & $-$38.523 &           & 56.4, 78.6 &     \\
       &     & 301.622  & $-$38.663 &           & 56.0, 81.2 &     \\
       &     & 301.302  & $-$37.024 &           & 56.6, 79.3 &     \\
060317 & 5.3 & 301.900  & $-$10.882 & (center)  & 125.0, 49.5 & 9.1  \\
       &     & 301.391  & $-$17.546 &           & 131.7, 49.5 &      \\
       &     & 301.976  & $-$8.163  &           & 122.3, 49.4 &      \\
       &     & 303.343  & $-$4.329  &           & 118.4, 50.7 &      \\
       &     & 301.987  & $-$13.575 &           & 127.7, 49.9 &      \\
060429 & 0.14& 115.314  & $-$24.926 & (center)  & 146.1, 134.1 & 45.5\\
       &     & 115.307  & $-$24.932 &           & 146.1, 134.1 &     \\
       &     & 113.108  & $-$25.520 &           & 145.5, 130.5 &     \\
       &     & 117.509  & $-$24.233 &           & 146.5, 137.8 &     \\
       &     & 115.321  & $-$24.919 &           & 146.0, 134.0 &     \\
060610 & 0.013& 354.762  &  52.107 & (center)  & 41.1, 23.6   &  48.3 \\
       &     & 354.294  &  52.375 &           & 40.7, 23.5 &       \\
       &     & 354.260  &  52.417 &           & 40.6, 23.5 &       \\
       &     & 355.227  &  51.835 &           & 41.4, 23.7 &       \\
       &     & 355.261  &  51.793 &           & 41.5, 23.7 &       \\
\hline 
\multicolumn{7}{l}{\scriptsize{$\ast$: $\theta$ and $\phi$ are the zenith and
  azimuthal angles, respectively, in degrees.}}\\
\end{tabularx}
\end{center}
\end{table}


\clearpage


\begin{table}[htbp]
\caption{Results of spectral fittings.}
\label{fit_summary_tbl}
\begin{tabularx}{17cm}{clcccccc}
\hline
\hline
\multicolumn{8}{c}{GRB 051127} \\
\hline
Model & Detector & $\alpha^\ast$ & $\beta^\dag$ & E$_{\rm peak}$ [keV]& Norm$^\ddag$ & Const$^\S$ &
 $\chi^2/ d.o.f$ \\
\hline
PL       & WAM0    & $-$1.24$^{+0.05}_{-0.05}$ & - & - &3.78$^{+0.41}_{-0.36}$& - & 142/24 \\
         &  KW(S1)  & $-$1.07$^{+0.07}_{-0.07}$ & - & - & 4.67$^{+0.58}_{-0.59}$ & - & 70/31 \\ 
         & WAM0+KW(S1) & $-$1.22$^{+0.05}_{-0.04}$ & - & - & 5.30$^{+0.48}_{-0.54}$ &
 0.67$^{+0.10}_{-0.07}$	 & 229/56\\
\hline
CPL           & WAM0   & $-$0.50$^{+0.18}_{-0.18}$ & - & 1257$^{+280}_{-192}$ &
  2.58$^{+0.40}_{-0.35}$& - & 29/23 \\
              & KW(S1) & $-$0.29$^{+0.29}_{-0.21}$ & -	& 940$^{+280}_{-181}$ &
 5.37$^{+0.87}_{-0.82}$  &	- &	32/30 \\ 
              & WAM0+KW(S1) &$-$0.44$^{+0.15}_{-0.15}$ & - & 1168$^{+207}_{-150}$& 5.18$^{+0.57}_{-0.57}$ & 0.50$^{+0.07}_{-0.06}$	& 64/55\\
\hline
Band          & WAM0  & $-$0.30$^{+0.18}_{-0.24}$ & $< -$2.08 &
 999$^{+347}_{-273}$ & 2.48$^{+0.40}_{-0.31}$ & - & 32/22 \\
             & KW(S1)  & $-$0.34$^{+0.19}_{-0.17}$ &	$< -$2.03 &
 999$^{+58.5}_{-187}$ & 5.30$^{+0.87}_{-0.83}$ & - & 32/29\\
           & WAM0+KW(S1)&$-$0.32$^{+0.13}_{-0.10}$& $< -$2.33 &
 1038$^{+185}_{-110}$ & 5.16$^{+0.58}_{-0.56}$ & 0.48$^{+0.06}_{-0.06}$ & 64/54 \\
\hline
\multicolumn{8}{c}{GRB 060317} \\
\hline
Model & Detector & $\alpha$ & $\beta$ & E$_{\rm peak}$ & Norm & Const &
 $\chi^2/ d.o.f$ \\
\hline
PL    &WAM0   & $-$1.11$^{+0.05}_{-0.06}$ & - & - &2.14$^{+0.48}_{-0.34}$  & - & 54/25 \\
      &KW(S2) & $-$1.32$^{+0.05}_{-0.05}$ & - & - & 0.90$^{+0.08}_{-0.08}$ & - & 74/50 \\ 
      & WAM0+KW(S2) & $-$1.16$^{+0.05}_{-0.05}$ & - & - &
 0.70$^{+0.08}_{-0.08}$ & 3.27$^{+0.25}_{-0.24}$	 & 167/76\\
\hline
 CPL   &WAM0& $-$1.03$^{+0.05}_{-0.06}$ & - & $>$5000 & 2.06$^{+0.48}_{-0.33}$  & - & 50/24 \\
       &KW(S2)&$-$1.02$^{+0.15}_{-0.13}$  & -	& 2089$^{+1200}_{-634}$ &
 0.87$^{+0.09}_{-0.09}$  &	- & 53/49 \\   
       & WAM0+KW(S2)& $-$1.03$^{+0.08}_{-0.09}$  & - & 5687$^{+3092}_{-1672}$ &
 0.70$^{+0.08}_{-0.08}$ & 3.07$^{+0.25}_{-0.22}$& 144/75\\
\hline
Band  &WAM0&$-$0.99$^{+0.11}_{-0.14}$  & $-$2.50$^{+0.06}_{-0.09}$ &
 $>$5000 & 2.02$^{+0.48}_{-0.33}$ & - & 47/23 \\
           & KW(S2)  & $-$0.85$^{+0.34}_{-0.06}$ &	$< -$2.73 &
 997$^{+1089}_{-1281}$ & 0.93$^{+0.08}_{-0.10}$ & - & 55/48\\
           & WAM0+KW(S2)&$-$0.99$^{+0.04}_{-0.15}$& $< -$1.58 &3230$^{+6743}_{-2232}$ & 0.76$^{+0.03}_{-0.14}$ & 2.74$^{+0.26}_{-0.19}$ & 128/74 \\
\hline
\multicolumn{8}{X}{\scriptsize $\ast$:power law photon index for PL and CPL
 model, and low energy index for the Band model.}\\
\multicolumn{8}{X}{\scriptsize $\dag$:power law photon index for the high energy
 component of the Band model.}\\
\multicolumn{8}{X}{\scriptsize $\ddag$:normalization in photons keV $^{-1}$ cm$^{-2}$ s$^{-1}$ at 100 keV.}\\
\multicolumn{8}{X}{\scriptsize $\S$:constant factor for the WAM to Konus
 normalization in joint spectral fittings.}
\end{tabularx}
\end{table}

\begin{table}[htbp]
\caption{(Continued.)}
\begin{tabularx}{17cm}{clcccccc}

\hline
\multicolumn{8}{c}{GRB 060429} \\
\hline
Model & Detector & $\alpha$ & $\beta$ & E$_{\rm peak}$ & Norm & Const &
 $\chi^2/ d.o.f$ \\
\hline
PL    &WAM0& $-$1.24 $^{+0.12}_{-0.10}$ & - & - & 3.66$^{+1.15}_{-0.78}$ & - & 42/25 \\
      &KW(S1) & $-$1.27$^{+0.04}_{-0.04}$ & - & - & 9.55$^{+0.85}_{-0.85}$ & - & 98/33 \\ 
      &WAM0+KW(S1) & $-$1.26$^{+0.05}_{-0.04}$ & - & - & 9.50$^{+0.88}_{-0.85}$ &
             0.40$^{+0.05}_{-0.05}$	 & 140/59\\
\hline
CPL   &WAM0& $-$0.99 $^{+0.26}_{-0.28}$ & - & $>$1430 & 3.29$^{+1.21}_{-0.83}$ & - & 37/24 \\
      & KW(S1)   & $-$0.76$^{+0.15}_{-0.13}$ & -	& 1583$^{+492}_{-371}$ &
   11.0 $^{+1.17}_{-1.14}$&	- &	30/32 \\   
      & WAM0+KW(S1)& $-$0.80$^{+0.12}_{-0.13}$ & - &  1720$^{+504}_{-384}$&
 10.9$^{+1.13}_{-1.06}$ & 0.28$^{+0.07}_{-0.05}$	& 70/58\\
\hline
 Band &WAM0 & $-$0.99$^{+0.55}_{-0.21}$ & $< -$1.28 &
 $>$1000 & 3.29$^{+0.78}_{-0.85}$ & - & 37/23 \\
      & KW(S1)  &$-$0.62 $^{+0.14}_{-0.29}$  &	$< -$2.02 &
 1560$^{+499}_{-656}$ & 11.0$^{+1.25}_{-1.09}$ & - & 30/31\\
      & WAM0+KW(S1)&$-$0.62$^{+0.12}_{-0.17}$& $< -$1.67 & 1251$^{+519}_{-581}$ &11.2$^{+0.97}_{-1.10}$ & 0.27$^{+0.07}_{-0.05}$ & 73/57 \\
\hline
\multicolumn{8}{c}{GRB 060610} \\
\hline
Model & Detector & $\alpha$ & $\beta$ & E$_{\rm peak}$ & Norm & Const & $\chi^2/ d.o.f$ \\
\hline
PL       & WAM1    & $-$1.16$^{+0.08}_{-0.09}$ & - & - & 3.46$^{+0.82}_{-0.73}$& - & 38/25 \\
         &KW(S2)   & $-$1.33$^{+0.04}_{-0.04}$ & - & - & 6.25$^{+0.48}_{-0.48}$ & - & 123/50 \\ 
         &WAM1+KW(S2)       &  $-$1.29$^{+0.04}_{-0.05}$ & - & - & 5.98 $^{+0.52}_{-0.53}$ &           0.66$^{+0.14}_{-0.12}$	 & 175/76\\
\hline
CPL        & WAM1   & $-$0.90$^{+0.17}_{-0.22}$ & - & 2889$^{+2973}_{-1175}$ &           3.11$^{+0.83}_{-0.74}$& - & 24/24 \\
           & KW(S2)   & $-$0.94$^{+0.12}_{-0.11}$ & -	& 1536$^{+500}_{-366}$ &           7.18$^{+0.66}_{-0.63}$  & - &67/49	 \\ 
           & WAM1+KW(S2)& $-$0.92$^{+0.09}_{-0.12}$ & - & 1821$^{+495}_{-364}$& 6.71$^{+0.63}_{-0.67}$ &           0.52$^{+0.11}_{-0.09}$	& 104/75\\
\hline
Band       & WAM1  & $-$0.81$^{+0.18}_{-0.23}$ & $<$ $-$8.82 &           1799$^{+1855}_{-817}$ & 3.12$^{+0.83}_{-0.73}$ & - & 24/23 \\
           & KW(S2)  & $-$0.99$^{+0.15}_{-0.04}$ &	$< -$2.21 &           1531$^{+528}_{-414}$ & 7.29$^{+0.58}_{-0.73}$ & - & 69/48\\
           & WAM1+KW(S2)&$-$0.88$^{+0.10}_{-0.13}$& $< -$2.50 & 1687$^{+634}_{-688}$ & 6.70$^{+0.65}_{-0.71}$ & 0.50$^{+0.16}_{-0.07}$ & 102/74 \\
\hline
\end{tabularx}
\end{table}

\begin{table}[htbp]
\begin{center}
\caption{Fluence and peak flux of our four short GRBs.}
\label{flue_and_pkflux}
\begin{tabularx}{18cm}{c|cccc|cc}
\hline
\multicolumn{1}{c}{GRB} & \multicolumn{4}{c}{Fluence$^\ast$} & \multicolumn{2}{c}{Peak
 flux$^\dag$}\\
\hline
    & \multicolumn{4}{c|}{[10$^{-6}$ erg cm$^{-2}$]}&[10$^{-6}$
 erg s$^{-1}$ cm$^{-2}$] & \multicolumn{1}{|c}{[ph s $^{-1}$ cm$^{-2}$]}\\
\multicolumn{1}{c|}{}&   50-100 keV & 100-300 keV & 300-500 keV &
 500-1000 keV &\multicolumn{1}{c}{} &\multicolumn{1}{|c}{}\\
\hline
GRB 051127 &  0.16$^{+0.02}_{-0.02}$ &
 0.93$^{+0.08}_{-0.09}$ & 1.1$^{+0.11}_{-0.11}$ &
 2.31$^{+0.24}_{-0.24}$ & 
 28$^{+3.2}_{-3.2}$ &
 22$^{+2.5}_{-2.5}$ \\
GRB 060317 & 0.72$^{+0.06}_{-0.06}$ &
 2.60$^{+0.02}_{-0.02}$ & 2.38$^{+0.02}_{-0.02}$ &
 5.47$^{+0.04}_{-0.04}$ & 
 46$^{+3.8}_{-3.7}$ & 41$^{+3.4}_{-3.3}$\\
GRB 060429 & 0.10$^{+0.01}_{-0.01}$ &
 0.49$^{+0.04}_{-0.04}$ &
 0.49$^{+0.05}_{-0.05}$ & 1.10$^{+0.12}_{-0.12}$ &
 67$^{+6.7}_{-5.7}$ &
 77$^{+7.7}_{-6.6}$ \\
GRB 060610 & 0.13$^{+0.01}_{-0.01}$ &
 0.55$^{+0.04}_{-0.04}$ &
 0.52$^{+0.05}_{-0.05}$ & 1.12$^{+0.11}_{-0.10}$ &
 33$^{+2.3}_{-2.3}$ &
 34$^{+2.3}_{-2.3}$ \\
\hline
\multicolumn{6}{X}{\scriptsize $\ast$: The energy fluence obtained by the CPL
 model in the joint fitting of the WAM data.} \\
\multicolumn{6}{X}{\scriptsize $\dag$: The energy and photon peak flux in
 1/32 s for GRB 051127 and 060317, and in 1/64 s for GRB 060429 and 060610, respectively, obtained by the WAM
 50-5000 keV data.} \\
\end{tabularx}
\end{center}
\end{table}

\clearpage
\begin{table}[htbp]
\caption{Duration(T$_{90}$) and spectral lag of four short GRBs and two
 long GRBs obtained by the WAM data. }
\label{temporal_summary_tbl}
\begin{center}
\begin{tabularx}{14cm}{ccccc}
\hline
GRB Date & T$_{90}$$^\ast$ & $\tau_{\rm lag}$ (TH0,TH1)$^\dag$  &  $\tau_{\rm lag}$ (TH0,TH2)$^\dag$ &  $\tau_{\rm lag}$ (TH0,TH3)$^\dag$ \\
         &     [s]          & \multicolumn{3}{c}{[ms]}\\
\hline
\multicolumn{5}{c}{short GRBs}\\
051127   & 0.66    & 17$^{+18}_{-21}$       & 13$^{+14}_{-14}$     & 7.1$^{+23}_{-18}$\\
060317   & 1.31     & 5.1$^{+5.7}_{-5.9}$        & 8.4$^{+5.6}_{-6.0}$      & 7.3$^{+8.4}_{-7.2}$ \\
060429   & 0.08    & 5.4$^{+6.6}_{-12}$        & 12$^{+14}_{-15}$      & 11$^{+26}_{-28}$ \\
060610   & 0.79     & $-$17$^{+33}_{-25}$       & $-$21$^{+36}_{-50}$      & 20$^{+36}_{-45}$ \\
\hline
\multicolumn{5}{c}{long GRBs}\\
050924   & 29.09   & 263$^{+83}_{-59}$        & 588$^{+89}_{-124}$      & 1255$^{+525}_{-266}$   \\
060915   & 43.16   & 56$^{+27}_{-18}$         & 131$^{+30}_{-30}$       & 170$^{+62}_{-72}$    \\
060922   & 13.88   & 199$^{+47}_{-31}$        & 505$^{+36}_{-56}$      & 1004$^{+116}_{-141}$    \\
\hline
\multicolumn{5}{X}{\scriptsize $\ast$: T$_{90}$ duration measured in 50-5000
 keV energy band of the WAM light curve.}\\
\multicolumn{5}{X}{\scriptsize $\dag$: The energy bands of TH0, TH1, TH2, and TH3 correspond to 50-110 keV,
 110-240 keV, 240-520 keV, and 520-5000 keV, respectively.}\\
\end{tabularx}
\end{center}
\end{table}
\vspace{-0.6cm}

\vspace{1cm}

\clearpage

\end{document}